\documentclass[10pt,twoside,leqno]{article}

\usepackage{a4wide} % for equation refing
\usepackage{amsmath} % for equation refing
\usepackage{amssymb} % math symbols
\usepackage{subfigure} % subfigures within figures
\usepackage{graphicx}
\usepackage{natbib}
\usepackage{natbibspacing} % following 2 lines alter spacing between
\usepackage{float}
\usepackage[small]{caption}
\usepackage{rotating}
\usepackage{booktabs}
\usepackage{url}
\usepackage{enumerate}
\usepackage{bm}
\usepackage{nopageno}
\usepackage{enumerate}

\floatstyle{ruled}
\newfloat{algorithm}{htbp}{lop}
\floatname{algorithm}{Box}

\newcommand{\tab}{\hspace*{5mm}} % tab

\begin{document}

	\bigskip

	\begin{center}

	{\Large \textbf{Fast Identification of Biological Pathways
		Associated with a Quantitative Trait Using
		Group Lasso with Overlaps}}

	\bigskip

	BY MATT SILVER, GIOVANNI MONTANA\footnote{Corresponding author. Email: {\tt g.montana@ic.ac.uk} }\\ \& ALZHEIMER'S DISEASE NEUROIMAGING INITIATIVE

    \bigskip

	Department of Mathematics\\
	Imperial College London\\London, SW7 2AZ, UK\\

	\end{center}
	
\begin{abstract}
	Where causal SNPs (single nucleotide polymorphisms) tend to accumulate within biological
	pathways, the incorporation of prior pathways information into a statistical model is expected
	to increase the power to detect true associations in a genetic association study. Most existing
	pathways-based methods rely on marginal SNP statistics and do not fully exploit the dependence
	patterns among SNPs within pathways.
	We use a sparse regression model, with SNPs grouped into pathways, to identify causal
	pathways associated with a quantitative trait. Notable features of our “pathways group lasso with
	adaptive weights” (P-GLAW) algorithm include the incorporation of all pathways in a single
	regression model, an adaptive pathway weighting procedure that accounts for factors biasing
	pathway selection, and the use of a bootstrap sampling procedure for the ranking of important
	pathways. P-GLAW takes account of the presence of overlapping pathways and uses a novel
	combination of techniques to optimise model estimation, making it fast to run, even on whole
	genome datasets.
	In a comparison study with an alternative pathways method based on univariate SNP statistics,
	our method demonstrates high sensitivity and specificity for the detection of important pathways,
	showing the greatest relative gains in performance where marginal SNP effect sizes are small.
\end{abstract}
	
\section{Introduction}

The mixed success of attempts to identify genetic variants that account for a large part of the heritability of common disease has focussed attention on the need to develop new methodological approaches to the analysis of GWAS data.  A number of factors that might explain this `missing heritability' have been suggested, including the failure of many current models to capture the presence of gene-gene and gene-environment interactions, of multiple SNPs with small effect and of rare variants \citep{Manolio2009,Goldstein2009}.  One promising approach uses prior information on functional structure present within the genome to group genes and associated SNPs into gene sets or pathways.  The motivation here is that genes do not work in isolation, but instead work together through their effect on molecular networks and cellular pathways.  The hope is that by jointly considering the effects of multiple SNPs or genes within a biological pathway, significant associations might be identified that would otherwise be missed when considering markers individually \citep{Wang2010}.  First developed in the context of gene expression studies \citep{Mootha2003}, pathways-based methods have more recently been extended to the analysis of GWAS data \citep{Holmans2009,Luo2010,LangoAllen2010,Lambert2010}.  This has led to the identification of putative causal pathways for a number of diseases including Parkinson's Disease \citep{Lesnick2007}, Crohn's Disease \citep{Wang2009} and rheumatoid arthritis \citep{Eleftherohorinou2011}.  As well as offering the potential for increased statistical power, pathways-based genetic association studies (PGAS) can aid the biological interpretation of results through the identification of causal pathways, and may also facilitate comparisons between studies genotyping different variants that nonetheless map to common pathways \citep{Ma2010,Cantor2010}.

The majority of existing PGAS methods begin with a univariate test of association, in which individual SNPs are scored according to their degree of association with disease status or a quantitative trait. Various techniques are then used to combine these univariate statistics into pathway scores.  For example, the GenGen method \citep{Wang2007} first ranks all genes according to the value of the highest-scoring SNP within 500kb of each gene.  Pathway significance is then assessed by determining the degree to which high-ranking genes are over-represented in a given gene set, in comparison with the genomic background.  The PLINK tookit \citep{Purcell2007} also features a `set-based test', in which pathway significance is measured by taking the average, marginal p-value of a pre-determined maximum number of `uncorrelated' SNPs within the pathway.  Here, uncorrelated SNPs are defined as those whose pairwise linkage disequilibrium (LD) is below a certain threshold value. As a final step, where more than one pathway is considered a correction for multiple testing is generally made.

In contrast to univariate, `one SNP at a time' methods, multivariate or multi-locus methods allow all SNPs to be considered in the model at the same time, which can aid the identification of weak signals while diminishing the importance of false ones. One such approach consists of fitting a penalised, multivariate regression model, in which a subset of SNPs is selected by imposing a penalty on some suitably selected norm of the regression coefficients, as in Lasso regression \citep{Tibshirani1996}. This approach has been shown to yield higher statistical power, compared to more common `mass univariate linear models', especially with multivariate and high-dimensional quantitative traits \citep{Vounou2010}.  Several other studies have demonstrated the advantages of this approach for the detection of genetic associations. For example, \cite{Wu2009} use penalized logistic regression to select SNPs in a case-control study, and analyse two-way and higher-order SNP-SNP interactions. \cite{Hoggart2008} propose a similar method for SNP selection in a Bayesian context. 

A number of penalized regression techniques that allow prior information on the relationship between SNP markers to be incorporated into the model selection process have recently been proposed. For example, \citet{Zhou2010} group SNPs into genes, and utilise a useful property of the group lasso \citep{Yuan2006} to aid the detection of rare variants within genes. The GRASS method (Chen et al., 2010) begins by characterising within-gene variation as `eigenSNPs', obtained by principal component analysis (PCA). A combination of lasso and ridge regression, followed by permutations is then used to measure significance for a single pathway.  Finally, \cite{Zhao2011} use a combination of PCA and lasso regression to identify a subset of genes within a candidate pathway, followed by permutations to measure pathway significance. Once again this method considers one pathway at a time.

The search for SNPs, or quantitative trait loci (QTL) influencing quantitative traits is gaining momentum as a potentially more powerful way to study the underlying causes of complex disease \citep{Plomin2009}.  In the emerging field of neuroimaging genetics for example, in which we have a particular interest, quantitative data in the form of MRI or PET scans serve as a type of intermediate phenotype in the study of complex disorders such as Alzheimer's Disease (AD) or schizophrenia \citep{Bigos2010}.  We use genotype data from the Alzheimers Disease Neuroimaging Initiative (ADNI) dataset in this analysis.

Our focus here is on the identification of biological pathways associated with a quantitative trait.  Our assumption is that where causal SNPs are enriched in a pathway, the use of a regression model that selects SNPs that are grouped into pathways will have increased power, compared to a more traditional approach in which SNPs are considered one at a time.  We also seek a true, multivariate model which includes all mapped pathways at the same time.  The hope is that this will confer some of the benefits, in terms of detecting weaker signals and diminishing false positives, described earlier.  To achieve these ends, we use a modified version of the group lasso (GL) with SNPs grouped into pathways, and develop a fast estimation algorithm applicable to the case of non-orthogonal groups.  In order to rank pathways, we use a bootstrap sampling procedure to rank pathways in decreasing order of importance.  We face a number of challenges in applying GL to SNP and pathway data for the identification of implicated pathways. These include the fact that pathways overlap, since many SNPs map to multiple pathways; the problem of selection bias, that is the tendency of the model to select pathways having specific statistical properties irrespective of their association with phenotype; and the sheer scale of SNP datasets, making efficient estimation a necessity. 

We have found that the issue of overlapping pathways receives surprisingly little attention in the PGAS literature, given that the presence of overlaps might be expected to have a significant impact on the results of any PGAS analysis.  For example, variation in the number and distribution of causal SNPs with respect to genes that overlap multiple pathways will affect the number of pathways defined to be `causal', and different PGAS methods will be affected by such variation in different ways.  Additionally, the inclusion of multiple pathways in a single GL regression model presents a particular problem, since GL in its original formulation will not select pathways in the manner that we would wish.  To account for this we employ a variable expansion procedure, originally proposed in the context of microarray data analysis by \cite{Jacob2009}, that ensures that overlapping SNPs enter the regression model separately, for each pathway that they map to.

A number of factors may bias PGAS results, exaggerating pathway significance and giving rise to inflated numbers of false positives.  Depending on the methods used, and the underlying disease-causing mechanism, such factors are likely to include pathway size (measured in number of SNPs and/or genes), and the extent and distribution of pathway LD.  Common strategies employed by existing methods to reduce this bias include the use of permutation (of genes or phenotypes), and dimensionality reduction techniques such as PCA \citep{Fridley2011,Wang2010}. We propose a procedure that reduces bias by adjusting pathway weightings in the regression model according to the empirical bias in pathway selection frequencies obtained by fitting the GL model with a null response.  

One potential drawback of using a regression model in the analysis of genetic data is the typically very large number of predictors (here SNPs) that must be analysed.  While the use of penalized regression techniques at least makes the problem tractable when the number of predictors vastly exceeds sample size, the very large matrix calculations required can still make model estimation computationally infeasible.  To address this, we combine a number of techniques that speed up the estimation process including the use of an `active set' of predictors, a Taylor approximation of the GL penalty and efficient computation of pathway block residuals. The final estimation algorithm, which we call `Pathways Group Lasso with Adaptive Weights' (P-GLAW), is sufficiently fast to obviate the need either to undertake a preliminary stage of dimensionality reduction, or to consider pathways individually.  

We evaluate our method's performance in a Monte Carlo (MC) simulation study, using real genetic and pathway data with quantitative phenotypes simulated under an additive genetic model. We consider a range of scenarios with different causal SNP distributions and effect sizes.  We feel the use of real genotype and pathway data is crucial, so as to capture the complex distributions of gene size and number within a pathway, together with SNP LD patterns and overlaps between pathways, all of which may have a significant effect on pathway ranking performance.  To our knowledge, this is the first such PGAS power study using GL with real SNP and pathway data.  The evaluation of GL pathway ranking performance however presents a number of challenges. Firstly, as described above, variation in the number of causal pathways due to overlaps must be taken into account when evaluating performance over multiple MC simulations.  Secondly, we require a means of evaluating the degree to which causal pathways are represented amongst the top ranks.  Thirdly, since GL performs variable selection, not all causal pathways may be ranked, and ranking performance measures must reflect this. To address these issues we devise a battery of measures that aim to capture different aspects of ranking performance. Finally, we compare our method's performance with another common PGAS method, derived from univariate SNP statistics.

The article is organised as follows. Section \ref{sec:Methods} describes the GL model; our strategy for dealing with overlapping pathways, model estimation and speed-ups; our proposed bias-adjusted pathway weighting update procedure; our strategy for ranking pathways using a resampling procedure, and our proposed ranking performance measures. In Section \ref{sec:SimulationStudy} we describe the real biological data sets used in the experiments, the SNP to pathway mapping process, and the simulation framework used to evaluate both methods under consideration.  The results from these simulation studies are provided in Section \ref{sec:Results}, and we conclude in Section \ref{sec:Discussion} with a discussion and final remarks.

%%%%%%%%%%%%%%%%%%%%%%%%%%%%%%%%%%%%%%%%%%%%%%%%%%%%%%%%%%%%%%%%%%%%%%%%%%%%%%%%%%%%%%%%%%%%%%%%%%%%%%%%%%%%%%%%%%%%%%%%%%%%%%%%%%%%%%%%%%%%
\section{Methods} \label{sec:Methods}
\subsection{The group lasso for pathway selection}{\label{subsec:GL}

We assume $N$ unrelated individuals genotyped at $P$ SNPs, each with a univariate quantitative trait $y_i$, for $i=1,\ldots, N$. For an individual $i$, we denote by $x_{ij}$ the minor allele count for SNP $j$, for $j=1, \ldots, P$, and arrange these counts in an $(N\times P)$ design matrix $\mathbf{X}$.  Quantitative phenotypes are arranged in an $(N \times 1)$ column vector $\boldsymbol{y}$, and will be treated as quantitative responses in a regression model. 

We initially consider the situation where SNPs are partitioned into $L$ mutually exclusive pathways, or groups. Each group $\mathcal{G}_l$, for $l=1, \ldots, L$, is a subset of $\{1,2,\ldots,P\}$ of cardinality $S_l$, containing the indices $l_1, l_2, \ldots, l_{S_l}$ of the SNPs that belong to it, such that $\mathcal{G}_l \cap \mathcal{G}_{l'} = \emptyset$ for any $l \ne l'$. We denote by $\mathcal{G} = \{1,\ldots,P\}$, the set of all SNP indices.  We denote by $\mathcal{S} \subset \{1, \ldots, P\}$ the subset of SNPs that are {\it causal}, that is the SNPs influencing $y$, and additionally denote the cardinality of $\mathcal{S}$ by $S$. Accordingly, we denote by $\mathcal{C} \subset \{1,2,\ldots,L\}$ the subset of causal pathways containing one or more SNPs in $\mathcal{S}$, having cardinality $|\mathcal{C}|$. We denote the complement of $\mathcal{C}$ by $\mathcal{C}'$.  We also assume that $|\mathcal{C}| \ll L$, so that only a small proportion of all pathways are causal. Finally, we assume that $y$ can be optimally predicted, in the least squares sense, by a linear combination of allele counts corresponding to SNPs in pathway $\mathcal{G}_l$, where $l$ belongs to the set $\mathcal{C}$. 

We denote the vector of SNP regression coefficients $\boldsymbol{\beta}=(\beta_1, \ldots, \beta_P) \in\mathbb{R}^P$, and the parameter vector corresponding to SNPs in pathway $\mathcal{G}_l$ only as $\boldsymbol{\beta}_l=(\beta_{l_1}, \ldots, \beta_{l_{S_l}}) \in \mathbb{R}^{S_l}$. Under these assumptions, one or more elements of each $\boldsymbol{\beta}_l$ for $l \in \mathcal{C}$ are expected to be non-zero, whereas all the regression coefficients associated with SNPs that do not belong to $\mathcal{C}$ will be zero, that is $\boldsymbol{\beta}_l = \mathbf{0}$ for $ l \in \mathcal{C'}$. For example, for a single causal pathway $\mathcal{G}_l$ with causal SNPs $\{a, b\}$ in $\mathcal{S}$, the sparsity pattern might look like
\[
	\boldsymbol{\beta} = \{
	\underbrace{(0,\ldots,0)}_{\mathcal{G}_1},\ldots,
	\underbrace{(0, \ldots, \beta_{l_a}, 0 \ldots, \beta_{l_{b}}, 0, \ldots, 0)}_{\mathcal{G}_l} ,\ldots,  
	\underbrace{(0, \ldots, 0)}_{\mathcal{G}_L} 
	  \}. 
\]

A suitable regression model that would enforce the assumed block sparsity pattern above is the group lasso (GL) \citep{Yuan2006}, in which estimates for $\boldsymbol{\beta}$ are obtained by minimising the penalised least squares function

\begin{equation}
	f(\boldsymbol{\beta}) = \frac{1}{2} ||\boldsymbol{y}-\mathbf{X}\boldsymbol{\beta}||_2^2 + \lambda \sum_{l=1}^{L}w_l||\boldsymbol{\beta}_l||_2
	\label{eq:GL}
\end{equation}
with respect to $\boldsymbol{\beta}$, where $||\cdot||_2$ denotes the $\ell_2$ (Euclidean) norm and $w_l$ is a pathway weighting factor for group $l$.  Sparsity at the pathway level is encouraged through the imposition of an $\ell_1$ lasso penalty on $||\boldsymbol{\beta}_l||_2$, which ensures that SNPs belonging to pathways not selected by the model have zero regression coefficients.  For selected pathways, i.e.~those with $\boldsymbol{\beta}_l \ne \mathbf{0}$, SNP coefficients tend to shrink, through the imposition of a ridge-type penalty on $||\bm{\beta}_l||_2$.  The degree of sparsity is controlled by the regularisation parameter, $\lambda$, such that the number of pathways selected by the model increases with decreasing $\lambda$.  For a given $\lambda$, the block sparsity pattern is determined both by the data ($\bm{y}$ and $\mathbf{X}$), and by the distribution of pathway weights, $\bm{w} = (w_1, \ldots, w_l)$, such that an increase in $w_l$ means that pathway $l$ is less likely to be selected, whereas a decrease in $w_l$ will have the opposite effect.

The GL optimisation problem associated with minimising the objective function \eqref{eq:GL} is convex, and can be solved using coordinate descent methods.  Problems arise however in the situation where pathways overlap, that is when a SNP is allowed to belong to more than one pathway, so that $\mathcal{G}_l \cap \mathcal{G}_{l'} \neq \emptyset$ for some $l \ne l'$. Firstly, where groups overlap, the penalty term in \eqref{eq:GL} is no longer separable into groups, since the same SNPs occur in multiple pathways, and convergence using coordinate descent is no longer guaranteed \citep{Tseng2009}.  Secondly, if we wish to be able to select pathways independently, GL is unable to do this. We illustrate this last point using a simple example in Fig.~\ref{fig:overlapping pathways} A, where we consider only three pathways, $\mathcal{G}_1, \mathcal{G}_2$ and $\mathcal{G}_3$, two of which overlap. As a consequence of this, pathway parameter vectors $\boldsymbol{\beta}_1$ and $\boldsymbol{\beta}_2$ also overlap, since they have a number of SNPs in common (shaded dark grey).  If a shared SNP is selected (i.e.~it has a non-zero coefficient), then both pathways to which it belongs ($\mathcal{G}_1$ and $\mathcal{G}_2$) are also selected, since their corresponding pathway parameter vectors have non-zero $\ell_2$ norms.  The GL regression model thus does not meet our requirements, since in order to be able to rank pathways in order of importance, we wish to be able to distinguish overlapping pathways and select them independently. Conversely, where shared SNPs have zero coefficients, for example in the case that $\mathcal{G}_1$ is not selected in the model, then these SNPs will have zero coefficients in each and every pathway to which they belong (here $\mathcal{G}_1$ and $\mathcal{G}_2$).  Hence SNPs retained in the model are necessarily drawn from the complement of a union of (unselected) pathways.  We instead require retained SNPs to be drawn from a union of (selected) pathways, so that a SNP driving selection in one pathway may still have a zero coefficient in another.

\bigskip \bigskip

\begin{figure}[h!]
	\begin{center}
	 \includegraphics[trim = 80mm 39mm 80mm 40mm, clip, scale=0.5]{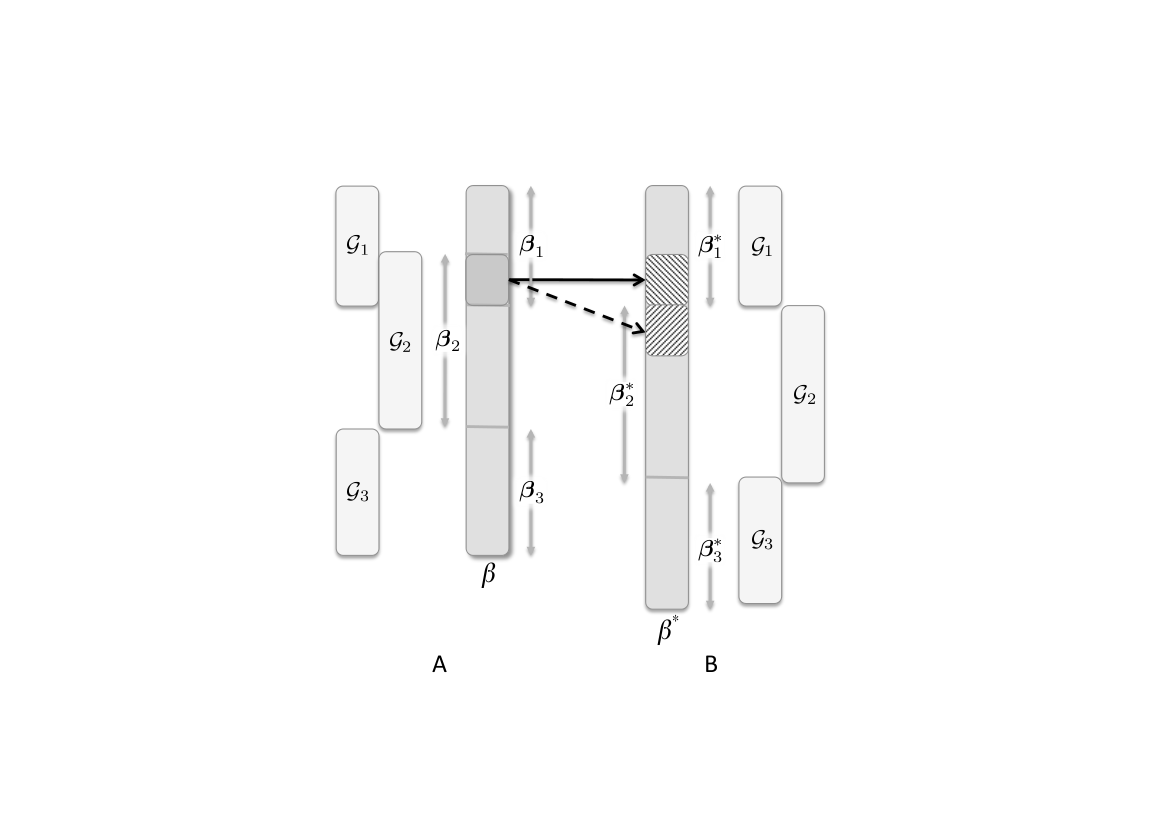}
	\caption{The problem of overlapping pathways: here there are three pathways, $\mathcal{G}_1, \mathcal{G}_2$ and $\mathcal{G}_3$, two of which overlap.  A: Standard formulation.  Pathway parameter vectors $\boldsymbol{\beta}_1$ and $\boldsymbol{\beta}_2$ overlap, since they have SNPs in common (shaded dark grey).  Where an overlapping SNP has a non-zero coefficient, only $\mathcal{G}_3$, can be selected independently.  B: Formulation with duplicated SNPs.  An expanded parameter vector, $\boldsymbol{\beta}^*$, is created by duplicating overlapping SNPs (dotted line).  $\boldsymbol{\beta}_1^*$ and $\boldsymbol{\beta}_2^*$ now enter the model separately, so that pathways can be selected independently.}
	\label{fig:overlapping pathways}
	\end{center}
\end{figure}

\cite{Jacob2009} propose one possible solution to the problem of overlapping predictors in a similar context, motivated by the analysis of gene expression data.  The essence of this method is to create duplicate, dummy SNPs, so that SNPs belonging to more than one pathway enter the model separately (see Fig.~\ref{fig:overlapping pathways} B).  The process works as follows.  An expanded design matrix is formed from the column-wise concatenation of the $L$ sub-matrices of size $(N \times S_l )$, that is 
$\mathbf{X}_{\mathcal{G}_l} = \{x_{ij}\}$ with $i = 1, \ldots, N$ and $j \in \mathcal{G}_l$, to form the expanded design matrix
$
	\mathbf{X}^* = [\mathbf{X}_{\mathcal{G}_1}, \mathbf{X}_{\mathcal{G}_2}, \dots , \mathbf{X}_{\mathcal{G}_L}]
$
 of size $(N \times P^*)$, where $P^* = \sum_l S_l$.  The corresponding parameter vector, $\boldsymbol{\beta}^*$, size $(P^* \times 1)$, is formed by joining the $L, ( S_l \times 1)$ pathway parameter vectors, $\boldsymbol{\beta}_l^*$, so that
$
	\boldsymbol{\beta}^* = [{\boldsymbol{\beta}_1^*}^T, {\boldsymbol{\beta}_2^*}^T, \dots,{\boldsymbol{\beta}_L^*}^T]^T
$.
The model is then able to perform pathway selection in the way that we require, and the optimisation \eqref{eq:GL}, with $\boldsymbol{\beta}$ replaced by $\boldsymbol{\beta}^*$, and $\mathbf{X}$ replaced by $\mathbf{X}^*$ becomes block separable, so that it can be solved by block coordinate descent.  In the following sections we assume both $\boldsymbol{\beta}$ and $\mathbf{X}$ have been expanded, but omit the $^*$ superscript for clarity. Finally, we note that where one or more SNPs in $\mathcal{S}$ overlap multiple pathways, the corresponding number, $|\mathcal{C}|$, of causal pathways will increase.

%%%%%% MODEL ESTIMATION

\subsection{Parameter estimation}\label{subsec:Model_estimation}

We seek a solution, $\boldsymbol{\hat{\beta}}$, that minimises the GL objective function \eqref{eq:GL}.  Where groups or pathways are disjoint, so that the penalties are separable into groups, a global solution can be obtained using block coordinate descent (BCD).   Coordinate descent algorithms offer a highly efficient means of solving convex optimisation problems, and work by breaking down the optimisation into a series of univariate problems, solving the optimisation for each variable (here SNP) in turn, while holding all the others fixed, until a suitable minimum based on some stopping criterion is reached \citep{Friedman2007}.  Where variables are grouped, as in GL, estimates are obtained for each pathway parameter vector, $\boldsymbol{\beta}_l$ in turn, while holding constant the current estimates for all other pathway parameter vectors, $\hat{\boldsymbol{\beta}}_m, (m \ne l)$, and then cycling through each pathway until convergence.

\cite{Yuan2006} derive a method for solving GL under the assumption that the group design matrices, $\mathbf{X}_{\mathcal{G}_l}$ are orthogonal, that is $\mathbf{X}_{\mathcal{G}_l}^T \mathbf{X}_{\mathcal{G}_l} = \mathbf{I} $.  This assumption does not hold in our case, so in the next section we derive a solution for GL in the case of non-orthogonal groups.  We additionally find that GL estimation using BCD can be slow, particularly for the large datasets common to PGAS, and so in the following sections propose a number of strategies for speeding up parameter estimation.
%%%%%%%%%%%%%%%%%%%%%%%%%%%%%%%%%
\subsubsection{Block coordinate descent for non-orthogonal groups} \label{subsec:bcd}

We assume that \eqref{eq:GL} is block-separable, that is the groups indexed by $1, \ldots, L$ are disjoint by construction. In our context, this is achieved by implementing the SNP duplication strategy of section \ref{subsec:GL}. We begin by considering a single pathway $l$. We collect the $N$ individual observed SNPs for a given SNP $j$ in a column vector $X_j = (x_{1j}, x_{2j}, \ldots, x_{Nj})$.  Using this notation, we define the matrix $\mathbf{X}_{\mathcal{G}_l} = (X_{l_1},X_{l_2},\dots, X_{S_l})$ containing all $S_l$ SNPs belonging to pathway $\mathcal{G}_l$, and the corresponding vector of regression coefficients $\boldsymbol{\beta}_l = (\beta_{l_1},\beta_{l_2},\dots,\beta_{S_l})$. We can then rewrite the objective function \eqref{eq:GL} for a single block $l$ as a function of $\boldsymbol{\beta}_l$,
\begin{equation}
	\label{eq:gl_singleGroup}
	 f(\boldsymbol{\beta}_l) = \frac{1}{2} || \hat{\mathbf{r}}_l - \sum_{j \in \mathcal{G}_l} X_j \beta_j  ||_2^2 + \lambda w_l ||\boldsymbol{\beta}_l||_2
\end{equation}
where $\hat{\mathbf{r}}_l = \mathbf{y} - \sum_{m \ne l} \mathbf{X}_m \boldsymbol{\hat{\beta}}_m$.  The vector $\hat{\mathbf{r}}_l$ is the `partial residual' vector for pathway $l$, based on the current estimates, $\boldsymbol{\hat{\beta}}_m, m \ne l$, of the other pathway parameter vectors.

Estimates for each $\beta_j$ are then obtained by taking partial derivatives with respect to $\beta_j$, that is by setting
\begin{equation}
	\frac{\partial f(\boldsymbol{\beta}_l)}{\partial \beta_j} = 0	\quad \text{for  } j = l_1,\dots,{S_l}
	\label{eq:obj_derivative}	
\end{equation}

Ignoring the penalty term, the partial derivative with respect to $\beta_j$ is
\[
	\frac{\partial}{\partial \beta_j} \frac{1}{2} || \hat{\mathbf{r}}_l - \sum_j X_j \beta_j  ||_2^2 \\
	= -X_j^T(\hat{\mathbf{r}}_l - \sum_j X_j \beta_j)
\]
We denote the partial derivative of the penalty term, by
\begin{equation*}
	s_j = \frac{\partial}{\partial \beta_j} || \boldsymbol{\beta}_l ||_2
\end{equation*}
so that \eqref{eq:obj_derivative} can be written as
\begin{equation}
	-X_j^T(\hat{\mathbf{r}}_l - \sum_j X_j \beta_j) + \lambda w_l s_j = 0
	\quad j = l_1,\dots,S_l
	\label{eq:diff_all_terms}
\end{equation}

We first consider the case where $\boldsymbol{\beta}_l = \mathbf{0}$, that is $\beta_j = 0$, for $j = l_1, \ldots, S_l$.  In this case $||\boldsymbol{\beta}_l||_2$ is not differentiable.  We instead form the $S_l$ sub-differentials, $s_j \in [-1,1]$, so that 
\begin{equation}
	\sum_j s_j^2  \le 1
	\label{eq:s_norm_0}
\end{equation}
The system of equations \eqref{eq:diff_all_terms} can now be written
\begin{equation*}
	s_j = \frac{1}{\lambda w_l} X_j^T \hat{\mathbf{r}}_l		\quad j = l_1,\ldots,S_l
\end{equation*}
and using \eqref{eq:s_norm_0}, we have
\begin{equation}
	\sum_j s_j^2 = \frac{1}{\lambda^2 w_l^2} \sum_j (X_j^T \hat{\mathbf{r}}_l)^2 \le 1.
	\label{eq:wl_size}
\end{equation}
Note that for \eqref{eq:wl_size} to be unbiased with respect to group size, a weight, $w_l = \sqrt{S_l}$, as proposed by \cite{Yuan2006}, can be applied.
Alternatively, since 
$$
\sum_j (X_j^T \hat{\mathbf{r}}_l)^2 = || \mathbf{X}_l^T \hat{\mathbf{r}}_l ||_2^2
$$
we may rewrite \eqref{eq:wl_size} as
\begin{equation*}
	(\sum_j s_j^2)^{\frac{1}{2}} = \frac{1}{\lambda w_l}	|| \mathbf{X}_l^T \hat{\mathbf{r}}_l ||_2 \le 1,
\end{equation*}
so that if $\boldsymbol{\beta}_l = \mathbf{0}$\\
\begin{equation}
	|| \mathbf{X}_l^T \hat{\mathbf{r}}_l ||_2 \le \lambda w_l.
	\label{eq:GL_theta_0}
\end{equation}\\

%% theta ne 0
When $\boldsymbol{\beta}_l \ne \mathbf{0}$, the minimisation of \eqref{eq:gl_singleGroup} can be obtained numerically, using coordinate descent, as a series of one-dimensional estimations over $\beta_j, j = l_1,\ldots,l_{S_l}$.  \cite{Friedman2010} suggest a golden section search over $\beta_j$, combined with parabolic interpolation. However, the number of such estimations depends on $L$ and $P^*$, both of which increase with $P$, the latter markedly so.  This can make the GL optimisation prohibitively slow, particularly for the large $P$ typically found in PGAS. For this reason, we next describe three strategies for speeding up the estimation.

%%%%%%%%%%%%%%%%%%%%%%%%%%%%%%%%%
\subsubsection{Taylor approximation of penalty} \label{subsec:taylorApprox}
One means of speeding up the estimation for $\beta_j$ is to use a linear or quadratic approximation of the GL $\ell_2$ penalty \citep{Zou2008,Fan2001}, enabling the replacement of the multi-step numerical optimisation over $\beta_j$ with a one-step calculation.  \cite{Breheny2009} propose the use of a Taylor approximation for a range of different estimation problems with grouped variables and we adopt this approach for our GL estimation problem.  We begin by rewriting the group $\mathcal{G}_l$ objective function \eqref{eq:gl_singleGroup}, for a single predictor as 
\begin{equation*}
	f(\boldsymbol{\beta}_l | \hat{\beta}_k, k \in \mathcal{G}_l, k \ne j) = 
		\frac{1}{2} || \hat{\mathbf{r}_l} - \sum_k X_k \hat{\beta}_k - X_j \beta_j  ||_2^2 + \lambda w_l \Gamma (\boldsymbol{\beta}_l | \hat{\beta}_k)
\end{equation*}
where $\Gamma (\boldsymbol{\beta}_l | \hat{\beta}_k) = (c + \beta_j^2)^{\frac{1}{2}}$, with $c = \sum_{k \ne j} \hat{\beta}_k^2$, and the $\hat{\beta}_k$ are the current SNP coefficient estimates.  For convenience, we rewrite this as
\begin{equation}
	f(\boldsymbol{\beta}_l | \hat{\beta}_k, k \ne j) = \frac{1}{2} || \hat{\mathbf{r}} + X_j \hat{\beta}_j - X_j \beta_j  ||_2^2 + \lambda w_l \Gamma (\boldsymbol{\beta}_l | \hat{\beta}_k)
	\label{eq:fl_theta}
\end{equation}
where $\hat{\mathbf{r}} = \mathbf{y} - \sum_{l} \mathbf{X}_l \boldsymbol{\hat{\beta}}_l$ is the total residual, using the current estimates of all SNP coefficients.  We now consider the first order Taylor expansion of $\Gamma (\boldsymbol{\beta}_l | \hat{\beta}_k)$ as a function of $x = \beta_j^2$, about the point $a = \hat{\beta}_j^2$\\
\begin{equation*}
	\Gamma(x) \simeq \Gamma(a) + \Gamma'(a)(x-a)
	\label{eq:GammaTaylor}
\end{equation*}
Now
\begin{align*}
	&\Gamma(x) = (c + x)^{\frac{1}{2}} \\
	\mbox{and }	&\Gamma'(a) = \frac{1}{2(c + a)^{\frac{1}{2}}}
\end{align*}
so that
\begin{equation*}
	\Gamma(x) \simeq (c + a) ^ {\frac{1}{2}} + \frac{x - a}{2(c + a)^{\frac{1}{2}}}
\end{equation*}\\\\
Substituting $a = \hat{\beta}_j^2$, and noting that $(c + a) ^ {\frac{1}{2}} = || \hat{\boldsymbol{\beta}_l} ||_2$, where $\hat{\boldsymbol{\beta}_l}$ denotes the current estimate of $\boldsymbol{\beta}_l$, this gives
\begin{equation*}
	\Gamma(\beta_j^2) \simeq \hat{\boldsymbol{\beta}_l} + \frac{\beta_j^2 - \hat{\beta}_j^2}{2|| \hat{\boldsymbol{\beta}_l} ||_2}
\end{equation*}
Substituting this expression in \eqref{eq:fl_theta}, we have
\begin{equation*}
		f(\boldsymbol{\beta}_l | \hat{\beta}_k, k \ne j) = \frac{1}{2} || \hat{\mathbf{r}} + X_j \hat{\beta}_j - X_j \beta_j  ||_2^2 + \lambda w_l 
		 	\Big[ \hat{\boldsymbol{\beta}_l} + \frac{\beta_j^2 - \hat{\beta}_j^2}{2|| \hat{\boldsymbol{\beta}_l} ||_2} \Big ]
\end{equation*}
Differentiating with respect to $\beta_j$ gives\\
\begin{align*}
	\frac{\partial f(\boldsymbol{\beta}_l)}{\partial \beta_j}\bigg|_{\hat{\beta}_k, k \ne j}
		&= -X_j^T (\hat{\mathbf{r}} + X_j \hat{\beta}_j - X_j \beta_j ) + \lambda w_l \frac{\beta_j}{|| \hat{\boldsymbol{\beta}_l} ||_2}\\
		&= -X_j^T \hat{\mathbf{r}} - \hat{\beta}_j + \beta_j + \lambda w_l \frac{\beta_j}{|| \hat{\boldsymbol{\beta}_l} ||_2}
\end{align*}
since $\sum_i x_{ij}^2 = X_j^T X_j =  1$.  Rearranging terms and setting the partial derivative equal to zero, we see that the minimum is achieved when
\begin{equation}
	\beta_j = \frac{X_j^T \hat{\mathbf{r}} + \hat{\beta}_j}{1 + \lambda'} \quad \mbox{where } \lambda' = \frac{\lambda w_l}{|| \hat{\boldsymbol{\beta}_l} ||_2}
	\label{eq:Taylor_thetaj}
\end{equation}
Where the current estimate $|| \hat{\boldsymbol{\beta}_l} ||_2 = \mathbf{0}$, that is when group $l$ first enters the estimation, we set $|| \hat{\boldsymbol{\beta}_l} ||_2 $ to be a small positive quantity, $\eta$, enabling $\beta_j$ in \eqref{eq:Taylor_thetaj} to be estimated.  

BCD proceeds by obtaining estimates for each $\beta_j, j = l_1, \ldots, S_l, 1,\ldots,$  $S_l, \ldots$ until convergence within the block, and for each  pathway, $l=1,\ldots,L,$ $1, \ldots, L, \ldots$ in turn, until a stopping criterion indicating a global minimum of \eqref{eq:GL} has been satisfied.  The estimation process is summarised in Box \ref{box:estimationAlgorithm}.

%%%%%%%%%%%%%%%%%%%%%%%%%%%%%%%%%
% BOX FOR BCD ALGORITHM
\begin{algorithm}[h]
\begin{enumerate}
\item
	set $\hat{\boldsymbol{\beta}} = \mathbf{0}$.  
	
	\item
	\begin{tabbing}
	kdk\=kdk\=kdk\=kdk\=kdk\=kdk\kill % set tabs
	For pathway $\mathcal{G}_l, l = l_1, 2, \ldots, L$: \\
	\>set $\hat{\mathbf{r}}_l = \mathbf{y} - \sum_{m \ne l} \mathbf{X}_m \boldsymbol{\hat{\beta}}_m$\\
	\>If $|| \mathbf{X}_l^T \hat{\mathbf{r}}_l ||_2 \le \lambda w_l$ \\
	\>\> set $\boldsymbol{\hat{\beta}}_l = \mathbf{0}$\\
	\>else\\
	\>\> do \>\\
	\>\>\> for $j = l_1,\dots,S_l$\\
	\>\>\>\>estimate $\beta_j$ using \eqref{eq:Taylor_thetaj}\\
	\>\>\>end\\
	\>\>until convergence of $f(\boldsymbol{\beta}_l)$ \eqref{eq:gl_singleGroup}\\
	\>\> set $\boldsymbol{\hat{\beta}}_l = \boldsymbol{\beta}_l$\\
	\>end\\
	end
\end{tabbing}
	\item
	Repeat step 2 until (global) convergence of $f(\boldsymbol{\beta}) \eqref{eq:GL}$
\end{enumerate}

\caption{GL estimation algorithm using BCD}
\label{box:estimationAlgorithm}
\end{algorithm}

%%%%%%%%%%%%%%%%%%%%%%%%%%%%%%%%%
\subsubsection{Use of pathway `active set'} \label{subsec:activeSet}
For large $P^*$ and $L$, the need for the repeated calculation of \eqref{eq:GL_theta_0} to establish whether or not a particular group can enter the estimation presents a major computational bottleneck.  This problem motivates another strategy providing substantial gains in computational efficiency for a range of sparse regression problems.  This `active set' strategy relies on the pre-selection of a subset of `potentially active' predictors, or groups of predictors that are likely to be selected by the model at a given $\lambda$ \citep{Tibshirani2010,Roth2008}.  The optimisation can then be run over this reduced set of variables, with a subsequent check to ensure that no other predictors should have been included in the first place.  The active set procedure offers potentially dramatic speed up in execution times, particularly for very large datasets such as those found in PGAS, due to the reduced number of computations that need to be performed.  In addition there are substantial savings in the amount of memory required to store data during processing, which can also lead to dramatic reductions in computation times with large datasets where memory is constrained.

For the GL, we begin by considering the inequality \eqref{eq:GL_theta_0}.  For groups to enter the model we require
\begin{equation}
	|| \mathbf{X}_l^T \hat{\mathbf{r}}_l ||_2 > \lambda w_l	\quad l = 1,\ldots,L
	\label{eq:kkt0r}	
\end{equation}
and therefore, at the first iteration, with $\boldsymbol{\beta}$ initialised to zero, a group $\mathcal{G}_l$ enters the model if
\begin{equation}
	|| \mathbf{X}_l^T \mathbf{y} ||_2 > \lambda w_l	\quad l = 1,\ldots,L.
	\label{eq:kkt0y}	
\end{equation}

We define the `active set' $\mathcal{A}$ of potentially active groups that satisfy \eqref{eq:kkt0y} as
\begin{equation*}
	\mathcal{A} = \{ m \in \mathcal{G} : || \mathbf{X}_m^T \mathbf{y} ||_2 > \lambda w_m \}	
\end{equation*}
and additionally define
\begin{equation}
	\lambda_{max} = \min_{\lambda}: || \mathbf{X}_l^T \mathbf{y} ||_2 \le \lambda w_l	\quad l = 1, \ldots, L
	\label{eq:lambdaMax}
\end{equation}
namely the smallest $\lambda$ value for which the active set is empty.  Note that provided $\lambda$ is close to $\lambda_{max}$, then $|\mathcal{A}| \ll L$.  Once one or more groups enter the model, not all $\hat{\boldsymbol{\beta}}_l$ will be zero and the inequality \eqref{eq:kkt0r} will then determine which groups may enter or leave the model.  

%%%%%%%%%%%%%%%%%%%%%%%%%%%%%%%%%
% BOX FOR ACTIVE SET ALGORITHM FOR A SINGLE LAMBDA
\begin{algorithm}[h]
\begin{enumerate}
\item
	Form the active set, $\mathcal{A} = \{ m \in \mathcal{G} : || \mathbf{X}_m^T \mathbf{y} ||_2 > \lambda w_m \}$
	\item
	Set $\hat{\boldsymbol{\beta}} = \mathbf{0}$, and solve the GL estimation at $\lambda$, using only the groups in $\mathcal{A}$:
	\begin{equation*}
	\hat{\boldsymbol{\beta}} = \min_{\boldsymbol{\beta}}  \frac{1}{2} ||\boldsymbol{y} - \sum_{m \in \mathcal{A}}
		\mathbf{X}_m \boldsymbol{\beta}_m||_2^2 + \lambda \sum_{m \in \mathcal{A}}w_m ||\boldsymbol{\beta}_m ||_2
	\end{equation*}
	\item
	Compute the revised active set on the full dataset:\\
		\begin{equation*}
			\mathcal{A}^+ = \{ z \in \mathcal{G} : || \mathbf{X}_z^T \hat{\mathbf{r}}_z ||_2 > \lambda w_z \}
		\end{equation*}
	if $\mathcal{A}^+ / \mathcal{A} = \emptyset$\\
		\tab $\hat{\boldsymbol{\beta}}$ is the full solution\\
		\tab STOP\\
	else\\
	\tab set $\mathcal{A} = \mathcal{A}^+$\\
	\tab repeat 2. and 3. with the new, (expanded) active set\\
	end
\end{enumerate}

\caption{Active set algorithm for a single $\lambda$ value}
\label{box:activeSetSingleLambda}
\end{algorithm}

%%%%%%%%%%%%%%%%%%%%%%%%%%%%%%%%%

The active set procedure rests on the observation that in practice, the final set of groups selected by the model rarely includes any groups not in $\mathcal{A}$ \citep{Tibshirani2010}.  We can therefore perform the full estimation on $\mathcal{A}$, followed by a check of the inequality \eqref{eq:kkt0r}, to see if any additional groups not in $\mathcal{A}$ can enter the model.  If there are no additional groups, then we have the full solution.  If not, then we run the full estimation again, with the additional groups satisfying \eqref{eq:kkt0r} added to $\mathcal{A}$.  A summary of the active set algorithm is given in Box \ref{box:activeSetSingleLambda}.

\subsubsection{Efficient computation of block residuals} \label{subsec:efficient_r_l}
A further, large computational burden results from the repeated calculation of the residuals $\textbf{r}_l$ and $\textbf{r}$ in \eqref{eq:GL_theta_0}, \eqref{eq:Taylor_thetaj} and \eqref{eq:kkt0r}.  The computational overhead for these calculations is substantial, both because of the size of the expanded design matrix ($N = 743$ and $P^*=66,085$ in the simulation study described in section \ref{sec:SimulationStudy}, but substantially larger for a full PGWAS), and because of the iterative nature of the BCD algorithm, meaning that a very large number of calculations are performed.  We therefore achieve one further substantial gain in computational efficiency by noting that since the blocks are separable, during BCD only the single block residual, $\mathbf{h}_l = \mathbf{y} - \mathbf{X}_l \boldsymbol{\beta}_l$, changes between iterations $j=1,\ldots,S_l,1,\dots,S_l,\dots$ within block $l$, and between iterations $l=1,\ldots,L,1,\dots,L,\ldots$ across blocks.  We therefore only need update $\mathbf{h}_l$ at each iteration, with $\mathbf{r}$ and $\mathbf{r}_l$ updated using computationally inexpensive matrix subtractions and additions.  Python code for mapping SNPs to pathways, and for analysing SNP data using PGLAW is available on request.

%\begin{figure}[h]
%\begin{center}
%	\includegraphics[scale=0.4]{GLexecutionTimes}
%\caption{Dependence of GL estimation time on the total number of SNPs, $P^*$, in $\mathbf{X}$, and effect of introducing more efficient estimation strategies described in the text.  `BCD' - estimation using block coordinate descent only.  `BCD+' - estimation using BCD with active set, Taylor approximation of the group penalty and efficient computation of block residuals.  Genotype and pathway data as described in section \ref{subsec:genotypePathwayData}, $N=743$.  Null $\mathcal{N} \sim (0,1)$ response.  
%$P^*= 4k: 5,000$ SNPs from chromosome 1 mapped to $126$ pathways.
%$P^*= 66k: 33,850$ genotyped SNPs from chromosome 1 mapped to $551$ pathways.
%$P^*= 647k: 448,294$ genome-wide SNPs mapped to $879$ pathways.  All computations performed using multi-threading on a single machine with $8$ $3.2$ GHz processors and $64$GB RAM.}
%\label{fig.executionTimes}
%\end{center}
%\end{figure}

%%%%%% PATHWAY WEIGHTING
\subsection{Selection bias and pathway weighting} \label{subsec:pathwayWeighting}

PGAS methods derived from univariate SNP statistics are subject to various biasing factors that can influence pathway ranking under the null, where no SNPs influence the phenotypic trait, $y$.  These factors vary from method to method, but may include the number and size of genes in a pathway, as well as LD between SNPs and genes. Such biasing factors are generally corrected through the use of permutation procedures.  For example, the `GenGen' method \citep{Wang2009}, measures the degree to which pathways are enriched with high ranking genes, and is subject to bias due to variation in the number of SNPs mapped to a gene, and to differences in LD between SNPs mapped to different genes.  The bias correction procedure begins by forming multiple datasets through permutation of phenotype labels.  For each permuted dataset, gene scores are generated from univariate SNP statistics, and a pathway enrichment score is calculated.  A normalised (bias-corrected) pathway enrichment score is then derived by comparing the distribution of pathway scores under the null with the score obtained from the unpermuted data.

Regression-based methods are similarly prone to bias, and once again the use of permutation has been proposed to correct for this, along with dimensionality reduction to extract non-redundant information.  For example, with the GRASS method for case-control data \citep{Chen2010}, genetic information within each gene is first summarised as `eigenSNPs', obtained through PCA.  The biasing effect of gene size is once again accounted for through the generation of a null distribution, formed by permuting phenotype labels.

With the GL under the null, pathway selection will be influenced by pathway size (i.e.~the number of SNPs within a pathway), since the accumulation of spurious associations in larger pathways will give rise to larger $|| \boldsymbol{\beta}_l ||_2$ in \eqref{eq:GL}.  In addition, variation in dependencies between SNPs within pathways, and to a lesser extent between pathways will give rise to corresponding variations in $|| \boldsymbol{\beta}_l ||_2$ where spurious associations arise in regions of high LD.

One way to correct for biases arising from variations in the statistical properties of different pathways or groups is through the selection of appropriate group weights $\bm{w}=(w_1, \ldots, w_L)$ for the objective function \eqref{eq:GL}. For example, as noted before, \cite{Yuan2006} suggest one possible choice for the pathway weighting would be
\begin{equation}
	w_l = \sqrt{S_l}
	\label{eq:std_size_weighting}
\end{equation}
which ensures that groups of different size are penalised equally, and so have an equal chance of being selected by the model, other things being equal (see \eqref{eq:wl_size}). In principle, we could follow this strategy and perhaps attempt to account for other, additional factors that may also bias pathway selection. However, there are a number of problems with this approach. Consider for example the biasing effect of dependencies between SNPs within a pathway.  Where causal SNPs tag, or reside within large blocks with strong LD, the pathway `signal' will be high, increasing the chance that such pathways will be selected by the model, compared with other pathways where LD is low.  This biasing effect will further depend on the distribution of LD within the pathway, which will in turn depend on other factors such as the number and size of pathway genes.  The precise form of any additional term(s) that should be added to \eqref{eq:std_size_weighting} to account for this bias is thus unclear.  Even if we were able to identify a list of potential biasing factors, and formulate bias-correcting weight adjustments for each, we are still faced with the problem that their may be other, unknown factors that contribute to the bias.  We therefore choose to adopt a `hypothesis-free' approach to adjusting pathway weights, which makes no assumptions about those factors which might influence pathway selection.  

Consider pathway selection under the GL model \eqref{eq:GL}, with $\lambda$ tuned to select $M$ pathways.  We begin with the case $M=1$. When there is no selection bias, and assuming no genetic association, a pathway $\mathcal{G}_l$ should be randomly selected by the model according to a uniform distribution, namely with probability $\Pi_l = 1/L$, for $l = 1,\ldots,L$. However, when biasing factors are present this is generally not the case, and the empirical probability distribution describing pathway selection, $\Pi^*(\bm{w})$ will not be uniform. Here the dependence upon the weight vector $\bm{w}$ has been made explicit, since with $\lambda$ tuned to select a single pathway, $\bm{w}$ alone determines the frequency distribution. %the observed distribution with GL selecting a single pathway under the null.  This depends on the data (genotypes and pathways), and is parameterised by $w$.
A measure of distance between these two distributions can be obtained by computing their Kullback-Leibler (KL) divergence
\begin{equation}
	D = \sum_l \Pi^*_l(\bm{w}) \log \frac{\Pi^*_l(\bm{w})}{\Pi_l} \label{eq:kldiv}
\end{equation}
where $\Pi^*_l(\bm{w})$ is the empirical probability for the selection of pathway $\mathcal{G}_l$ under the assumption of no genetic associations. When GL pathway selection is unbiased, we expect this distance to be approximately zero. Our strategy consists in adaptively adjusting all weights $\bm{w}$ in order to minimise $D$. 

%Since the empirical probability distribution $\Pi_l^*(w_l)$ describes pathway selection under the assumption of no genetic associations, this distribution can be generated by fitting the GL model over multiple datasets with permuted phenotype labels.  

%The weight adjustment under the null cannot however be accomplished in a single step for two reasons.  First of all, the presence of overlapping pathways means there is no analytical solution for finding the optimum $w$.  Secondly, in practice we find that pathway selection bias is initially high (Fig.~\ref{fig:adaptiveWeights} (a)), with many pathways having $\Pi^* = 0$, so that the weighting adjustment procedure must proceed iteratively.

Our adaptive weighting procedure is an iterative one, whereby at each iteration $\tau$ we first update the previous weight vector $\bm{w}^{(\tau-1)}$, and then re-estimate $\Pi^*(\bm{w}^{(\tau)})$ by fitting the GL model $R$ times, each with a random permutation of the response in order to create $R$ null data sets\footnote{Alternatively, in a simulation study where the null distribution of the response is known (as in section \ref{sec:SimulationStudy}), the $R$ models can be fitted after sampling a response from that null distribution.}. $\Pi^*_l(\bm{w}^{(\tau)})$ is then the frequency at which pathway $\mathcal{G}_l$ is selected across the $R$ null data sets at iteration $\tau$. The algorithm is initialised at iteration $\tau=0$ by using an initial weight vector $\bm{w}^{(0)}$, for instance the standard size weighting \eqref{eq:std_size_weighting}.   This procedure is then repeated until $D$ reaches some suitably small value.

From \eqref{eq:kldiv}, a reduction in $D$ can be obtained by reducing the difference $d_l = \Pi^*_l(\bm{w}) - \Pi_l$, for all $l$. As each $|d_l|$ approaches zero, the ratio, $\Pi^*_l(\bm{w})/\Pi_l$, approaches one, so that the contribution of pathway $\mathcal{G}_l$ to $D$ is decreased.  
With this in mind, at each iteration, we adjust pathway weights according to the following formula, 
\begin{equation}
	w_l^{(\tau)} 	= w_l^{(\tau-1)} \left[1 - \text{sign} (d_l) (\alpha - 1) L^2 d_l^2 \right] 	\qquad 0 <  \alpha < 1
	\label{eq:adjustWeights}	
\end{equation}
where the paramater $\alpha$ controls the maximum amount by which each $w_l$ can be reduced in a single iteration, in the case that pathway $\mathcal{G}_l$ is selected with zero frequency.  The weighting update equation has the following desirable properties.  When $0 \le \Pi^*_l < \Pi_l$, i.e.~$-\frac{1}{L} \le d_l < 0$, $w_l$ is decreased, up to a maximum factor $\alpha$ when $\Pi_l^* = 0$, increasing the chance that group $l$ is selected.  When $\Pi^*_l > \Pi_l$, i.e.~$d_l > 0$, $w_l$ is increased, decreasing the chance that group $l$ is selected.  Finally, when $\Pi^*_l = \Pi_l$, i.e.~$d_l = 0$, $w_l$ is unchanged.  The square in the weight adjustment factor ensures that large values of $|d_l|$ result in relatively large adjustments to $w_l$.  

The estimation of $\Pi^*$ when $M>1$, that is where more than one pathway is selected by the model, is computationally infeasible even for a small value of $M$, since we would need to estimate the empirical {\it joint} probability distribution that $M$ pathways are jointly selected. However, we expect that many of the factors biasing pathway selection when $M=1$ will similarly affect this joint probability distribution. Under this assumption, we estimate the optimal weight vector $\bm{w}$ only in the $M=1$ case. Extensive simulation studies (see section \ref{sec:Results}) indicate that this data-driven adaptive waiting scheme is able to substantially increase power and specificity compared with the standard weighting \eqref{eq:std_size_weighting}, even when $M>1$, indicating that this assumption holds in practice. Finally, we note that despite the need for multiple MC simulations over multiple iterations, our proposed bias-adjusted weighting strategy is fast, since it relies on fitting the GL model with $\lambda$ tuned to select a single pathway only, ensuring that the active set (see section \ref{subsec:activeSet}) is very small, and model estimation time for each of the $R$ model fits is minimal.

%    The assumption that pathways are selected independently, implicit in the previous $M=1$ case, now no longer holds.  For example, pathways with a high degree of overlap are more likely to be selected together.  Indeed we wish to exploit this feature of GL to select similar pathways in the situation where the response $y$ is influenced by a combination of SNPs that overlap multiple pathways.  The unbiased distribution $\Pi_M, (M>1)$, which describes selection frequency distribution when choosing $M$ pathways using GL under the null, is thus not uniform, i.e.~$\Pi_M(l) \ne 1/L, (M >1; l = 1, \ldots, L)$.  Even if we knew $\Pi_M$, its size grows very rapidly with $M$, making any study of the empirically observed distribution, $\Pi^*_M$, computationally infeasible.  

%%%%%% PATHWAY RANKING
\subsection{Pathway ranking}\label{subsec:PathwayRanking}

Penalized regression typically proceeds by determining an optimal value for $\lambda$, corresponding to a subset of variables that best predicts the response, and this is generally done by cross validating the prediction error.  In genetic association mapping, results are often instead presented in the form of lists of pathways or SNPs, ranked in order of importance.  We seek such a strategy for the ranking of pathways within the regression model, such that pathways in $\mathcal{C}$, will achieve a high ranking, whereas those in $\mathcal{C}'$ will be ranked low.  This approach has the added advantage of allowing us to make direct comparisons with alternative pathway methods that use p-values as a ranking criterion.

One simple ranking criterion in penalised regression is to use the order in which each variable enters the model along the regularization path - i.e.~as $\lambda$ is decreased from its maximal value, where no variables are selected.  We instead adopt a bootstrap sampling approach, in which we fit the regression model over multiple subsamples of the data, drawn with replacement, at a  single, \emph{fixed} value for $\lambda$.  Pathways are ranked in order of importance according to their selection frequency across subsamples.  Our motivation here is to exploit knowledge of finite sample variability obtained by subsampling, to achieve better estimates of pathway importance.  In this respect our strategy resembles the pointwise stability selection method proposed by \cite{Meinshausen2010} in the context of variable selection.

As with stability selection, for our ranking strategy to be effective, the value of $\lambda$ must be small enough to ensure that all pathways in $\mathcal{C}$ are selected by the model with a high probability at each subsample.  Computation time increases rapidly with $M$, the number of selected pathways, so that with the number, $|\mathcal{C}|$, of causal pathways unknown, the choice of $M$ is driven by the number of causal pathways we seek to identify within computational constraints.  We use $B = 100$ subsamples, each of size $N/2$, and at each subsample we perform a line search over $\lambda$, to ensure that $M \ge M_{min}$ pathways are selected.  This procedure is described in appendix \ref{appendix:lambdaLineSearch}.  Once $\lambda$ is tuned, for each subsample, $b$, we obtain estimates $\beta_{j}^{(b)} (b = 1, \ldots, B)$ for each SNP coefficient $(j = 1, \ldots, P^*)$.  For pathway $\mathcal{G}_l$, we define $\pi_l^{(b)} = 1$ when $||\boldsymbol{\beta}^{(b)}_l||_2 \ne 0$ and $\pi_l^{(b)} = 0$ otherwise, where $\boldsymbol{\beta}^{(b)}_l$ is the pathway parameter vector estimated for subsample $b$.  We rank pathways in order of their selection frequency across subsamples, $\bar{\pi}_{l_1} \ge, \ldots, \ge \bar{\pi}_{l_L}$. We note that since typically $M \ll L$, some $\bar{\pi}_{l}$ may be zero.  Such pathways are classified as unranked.

\subsection{Ranking performance measures}\label{subsec:RankingPerformanceMeasures}
In order to evaluate the success of any PGAS method, some measure of ranking performance is required.  In this section we describe 3 separate ranking performance measures that we use to evaluate the performance of our method in a simulation study described in section \ref{sec:SimulationStudy}.  One complicating factor is the issue of overlapping pathways, making the effective number of causal pathways, $|\mathcal{C}|$, dependent on the degree to which SNPs in $\mathcal{S}$ overlap multiple pathways.  In addition, with any method based on variable selection, the possibility that causal pathways are unranked, i.e.~they are not selected by the model, must be taken into account.  

Consider the situation where the set $\mathcal{S}$ of causal SNPs, with cardinality $S > 1$, is known.  We may choose to define $\mathcal{C}$ in its most restricted sense as the set of pathways that contain \emph{all} members of $\mathcal{S}$, or alternatively $\mathcal{C}$ might include all pathways containing one or more SNPs belonging to $\mathcal{S}$.  In either case $|\mathcal{C}|$ will depend on the degree to which SNPs in $\mathcal{S}$ overlap multiple pathways.  This in turn depends on the particular distribution of causal SNPs with respect to overlapping genes.  The need to accommodate this variability in $|\mathcal{C}|$ in part motivates our formulation of the ranking measures described below.

We propose three separate ranking measures that capture different aspects of ranking performance, and focus on the top 100 ranked pathways only.  We do this firstly because in any method attention is inevitably focused on the highest ranking pathways (or alternatively those with the highest statistical significance in a hypothesis testing framework).  Also, since in a simulation study we compare the performance of our variable selection method which identifies a limited number of pathways against an alternative method that scores all pathways, some suitable cutoff in rank order must be chosen.  

We denote the set of \emph{ranked} causal pathways by $\mathcal{C}^* = \{k \in \mathcal{C}: \bar{\pi}_{k} > 0 \}$, cardinality $|\mathcal{C}^*|$, and their respective rankings by $r_{k_1}, r_{k_2}, \ldots, r_{|\mathcal{C}^*|}$, ranked in order of their respective selection frequencies, $\bar{\pi}_{k_1} < \bar{\pi}_{k_2} < , \ldots, < \bar{\pi}_{|\mathcal{C}^*|}$.  We further denote by $\mathcal{C}^*_{100} = \{k \in \mathcal{C}^*: r_k \le 100 \}$, cardinality $|\mathcal{C}^*_{100}|$, the set of ranked causal pathways falling in the top 100 ranks, with corresponding rankings $r_{k_1}, r_{k_2}, \ldots, r_{|\mathcal{C}^*_{100}|}$.  Our three proposed ranking measures are as follows:

\begin{enumerate}
	\item
	\emph{Highest causal pathway rank, $r_{k_1}$}, that is the single highest rank achieved by any pathway in $\mathcal{C}^*_{100}$.  This lies in the range $1 \le r_{k_1} \le 100$, and is only defined for $|\mathcal{C}^*_{100}| \ge 1$.
	\item
	\emph{Ranking power, $p_{100}$}, defined as
\begin{equation}
	p_{100} = \frac{|\mathcal{C}^*_{100}|}{|\mathcal{C}|}
	\label{eq:rankingPower}
\end{equation} 
with $0 \le p_{100} \le 1$.  $p_{100} = 0$ when no causal pathways are ranked in the top 100 ($\mathcal{C}^*_{100} = \emptyset$), and $p_{100} = 1$ when all causal pathways are ranked in the top 100 ($\mathcal{C}^*_{100} = \mathcal{C}$).
	\item
	\emph{Power-adjusted, normalised, weighted ranking score, $R$}.  This takes account of the actual rankings, $r_{k_1}, \ldots, r_{|\mathcal{C}^*_{100}|}$, as well as the ranking power, $p_{100}$.  We begin by defining a normalised, weighted ranking score,
\begin{equation}
	R^* = \frac{\sum_{k \in \mathcal{C}^*_{100}} r_k ^ {\frac{1}{2}} } {\sum_{k = 1}^{|\mathcal{C}^*_{100}|} k ^{\frac{1}{2}}	}
	\label{eq:RankingScore}
\end{equation}
Here the square root increases the weight given to highly-ranked causal pathways.  The denominator is a normalising factor
which represents the minimum possible weighted ranking score, with $r_{k_1} = 1, r_{k_2} = 2 \ldots,$ $r_{|\mathcal{C}^*_{100}|} = |\mathcal{C}^*_{100}|$, ensuring that $R^*$ attains its minimum value of $1$ when the pathways in $\mathcal{C}^*_{100}$ are optimally ranked.  Higher values of $R^*$ indicate suboptimal ranking.  $R^*$ takes no account of the possibility that $\mathcal{C}^*_{100} \ne \mathcal{C}$, i.e.~not all causal pathways are ranked.  To do this we form the adjusted measure
\begin{equation}
	R = 
	\begin{cases}
		R^* / p_{100} 	&\mbox{if } p_{100}>0\\
		\gamma	&\mbox{if } p_{100}=0
	\end{cases}
	\label{eq:R}
\end{equation}
$R$ thus attains a minimum value of $1$ when all causal pathways are optimally ranked, and the value $\gamma$ when no causal pathways are ranked.
\end{enumerate}¥

%%%%%%%%%%%%%%%%%%%%%%%%%%%%%%%%%%%%%%%%%%%%%%%%%%%%%%%%%%%%%%%%%%%%%%%%%%
\section{Simulation Study}
\label{sec:SimulationStudy}

We assess the power of our proposed method in a simulation study using real genotype and pathway data, with simulated, quantitative phenotypes generated under an additive genetic model from SNPs within a single, selected causal pathway.  The presence of overlapping SNPs means that the actual number of causal pathways is typically greater than one.  We additionally compare our method's performance with an alternative, univariate-based method commonly used in gene set analysis.  Computation times for both methods increase with $P$, and because of this, and the large number of scenarios and simulations tested, we restrict this analysis to SNPs on a single chromosome to keep execution times within practical limits.  

%%%%%% GENOTYPES
\subsection{Genotype and pathways data} \label{subsec:genotypePathwayData}
We use genotypes obtained from the Alzheimer's Disease Neuroimaging Initiative, ADNI (\url{www.loni.ucla.edu/ADNI}), derived from the Illumina Human 610-Quad BeadChip.  Subjects comprise a mix of healthy controls, those diagnosed as having mild cognitive impairment, and those with AD.  After removing variants with a call rate $<95\%$, minor allele frequency (MAF) $<0.1$ and significant deviation from Hardy-Weinberg equilibrium ($p<5.7\times10^{-7}$), $448,294$ SNPs remain.  In this study we use genotype data from $N=743$ subjects, and consider only SNPs from chromosome $1~(33,850$ SNPs).

Popular databases used for the mapping of genes to biological pathways include the Kyoto Encylopedia of Genes and Genomes (KEGG, \url{www.genome.jp/kegg/pathway.html}) and BioCarta (\url{www.biocarta.com/genes/ index.asp}).  For this study we use data on `canonical pathways' from the Molecular Signals Database  (MSigDB, \url{www.broadinstitute.org/gsea/msigdb/index.jsp}), which is a commonly-used, curated collection of pathways obtained from multiple sources.  At the time of writing this comprised $880$ pathways mapped to $6,804$ genes.  2,382 human gene locations on chromosome 1, corresponding to assembly GRCh37.p3 are obtained using Ensembl's biomart API (\url{www.biomart.org}).  ADNI-genotyped SNPs on chromosome 1 are then mapped to annotated genes within 10kb (20,399 SNPs mapped to 2,096 genes), and these remaining genes and SNPs are then mapped to pathways using MSigDB (8,102 SNPs mapped to 778 pathways).  Thus we see that the majority of chromosome 1 SNPs fail to map to any pathway, but that the majority of annotated pathways map to at least 1 SNP on this chromosome.  Finally, small ($<10$ SNPs) and identical pathways are removed.  After all pre-processing we are left with a total of $P = 8,078$ SNPs mapped to $551$ pathways (max: $1,059$; min: $10$; mean: $120\pm142$ SNPs per pathway).  All SNP to pathway mapping and filtering was performed using bespoke code written in Python.  The mapping and filtering process is illustrated in Fig.~\ref{fig:SNPtoPathMapping}.

More than 80\% of SNPs are observed to overlap more than 1 pathway, with around 20\% overlapping 10 or more pathways and 2\% overlapping 60 or more (see Fig.~\ref{fig:paths_per_snp}).  After variable expansion to account for overlapping pathways (section \ref{subsec:GL}), we have $P^* = 66,085$ SNPs.  

\begin{figure}[h]
	\begin{center}
	\includegraphics[trim=7cm 1cm 6cm 1cm, clip, scale = 0.6]{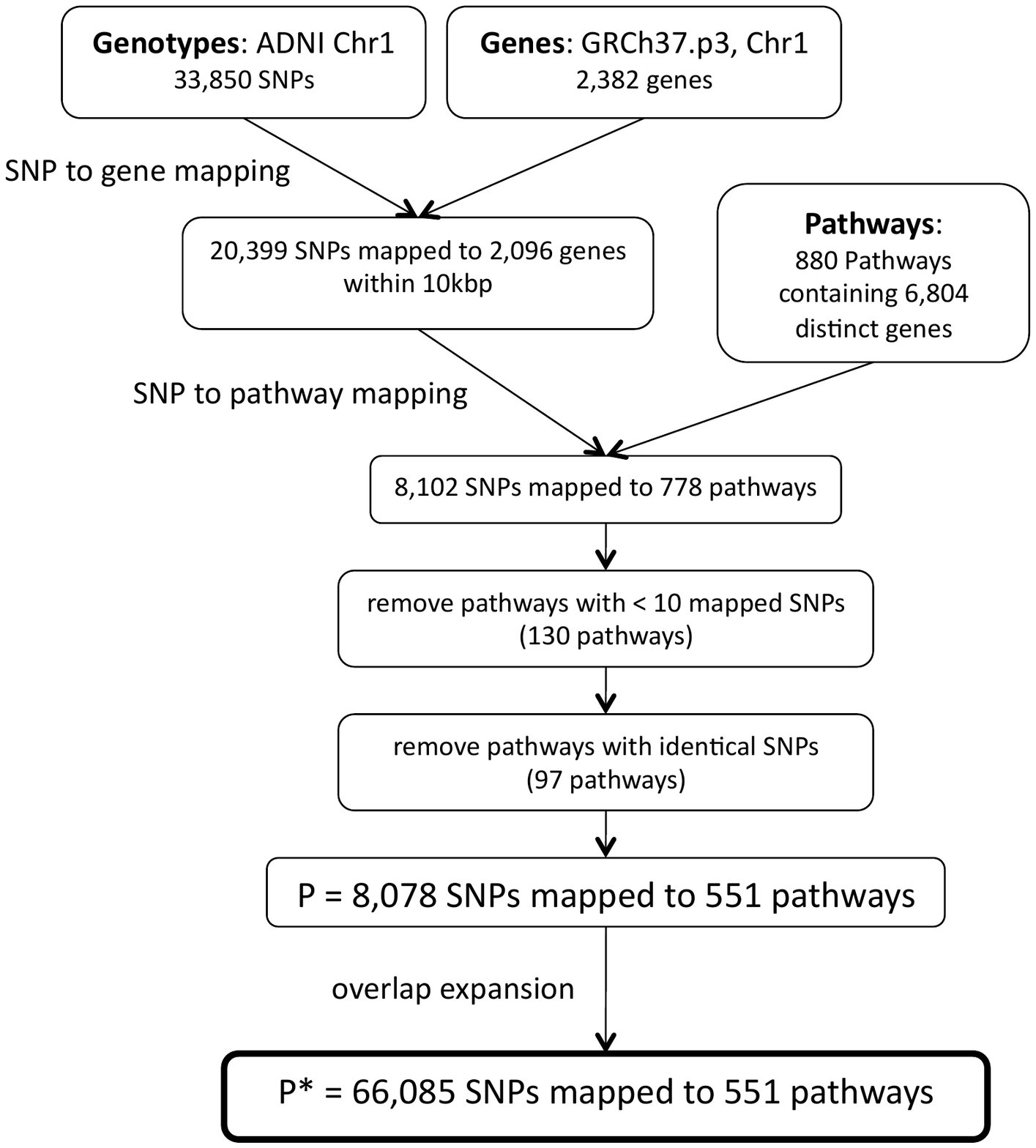}
	\caption{SNP to pathway mapping.}
	\label{fig:SNPtoPathMapping}
	\end{center}
\end{figure}

\begin{figure}
	\begin{center}
	\includegraphics[scale=0.5]{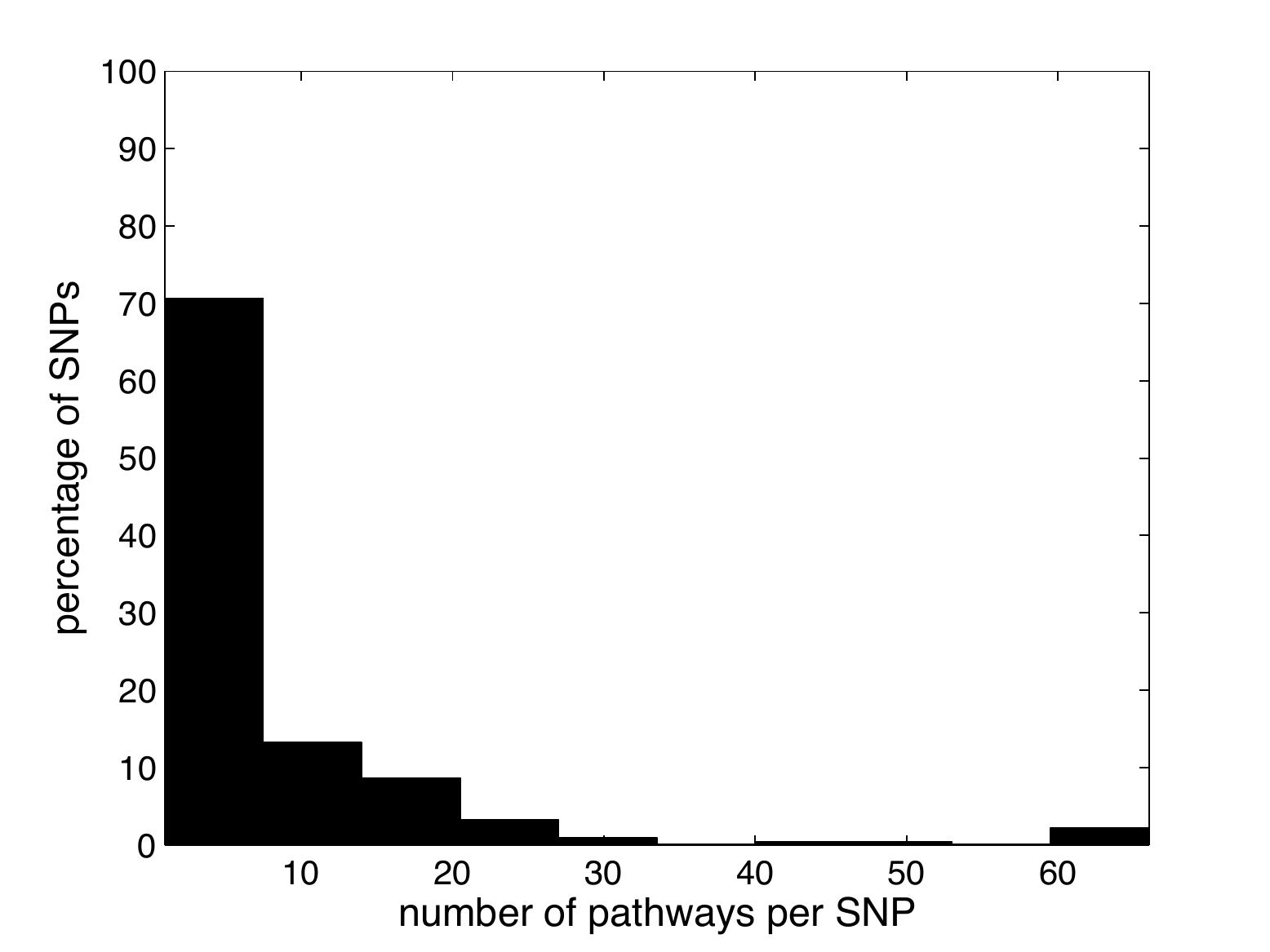}
	\caption{Frequency distribution of ADNI SNPs by number of pathways they map to. SNPs are mapped to genes within 10kbp. The data set consists of $8,078$ SNPs and $551$ pathways.}
	\label{fig:paths_per_snp}
	\end{center}
\end{figure}

\subsection{Simulation framework} \label{subsec:SimulationFramework}

We begin by adjusting the pathway weight vector, $\bm{w}$, using the bias-adjusted adaptive weighting procedure described in section \ref{subsec:pathwayWeighting}.  We do this over $10$ iterations with $R = 40,000$ MC simulations, each with response $y$ sampled from a standard normal distribution, $\mathcal{N}(0,1)$ for simplicity, since many quantitative traits are expected to follow a normal distribution. 

For the simulation of a SNP-dependent response, we begin by drawing $S$ SNPs from a single, randomly selected causal pathway, $\mathcal{G}_{\phi}$, according to some specified distribution (see below), and then form the set $\mathcal{C}$, of causal pathways that contain all the members of $\mathcal{S}$.  We thus chose to define $\mathcal{C}$ in its most restricted sense, rather than for example including pathways that contain one or more SNPs in $\mathcal{S}$.  Note that the number, $|\mathcal{C}|$ of causal pathways will vary according to the particular distribution of overlaps within $\mathcal{S}$.  

For each simulation, a univariate quantitative phenotype $y$ is simulated using an additive model, 

\begin{equation*}
	y_i = \sum_{k \in \mathcal{S} } \zeta_k x_{ik} + \epsilon	
\end{equation*}
where $\zeta_k$ is the allelic effect per minor allele due to causal SNP $k$.  Setting $w_k = \zeta_k x_k$, we define the \emph{effect size} of SNP $k$ as $\delta_k = \text{E}(w_k)/ \text{E}(y)$ for $k \in \mathcal{S}$, and set $\epsilon \sim \mathcal{N} (1, \sigma_{\epsilon}^2)$ so that $\delta_k = 0$ when $\zeta = 0$.  We also record the average SNP effect size as a proportion of total phenotypic variance, $\text{ES}_k = \text{Var}(w_k)/ \text{Var}(y)$, and the mean proportionate change in response per minor allele, $\text{E}(\zeta_k)$.  For our simulations we control $\delta_k$, and set $\zeta_k$ accordingly, so that effect size is independent of SNP MAF, whereas $\zeta_k$ and $\text{ES}_k$ are MAF-dependent.  

The power and specificity of any PGAS method is likely to depend on a range of factors including the number of causal pathways to be identified, the number and distribution of causal SNPs, and the size of their phenotypic effect \citep{Wang2010,Fridley2011}.  We therefore assess the performance of our method across $6$ different scenarios in which we vary each of these factors.  Furthermore, we test each scenario over $500$ MC simulations to account for variation in causal SNP MAFs, gene size and number within pathways, and  LD patterns within and between causal pathways.\\  

\begin{table}[ht]
	\begin{center}
	\footnotesize
	\tabcolsep 5pt
	\begin{tabular}{@{}c||ccll@{}}
	\hline
	scenario		& $S$	&	$\delta_k$	&	distribution	&	description\\
	\hline	
	(a)			&	$10$			&	$0.005$		&	random from $\mathcal{G}_{\phi}$
												&	$S$ large; $\delta_k$ large; random distribn\\
	(b)			&	$3$			&	$0.005$		&	random from $\mathcal{G}_{\phi}$
												&	$S$ small; $\delta_k$ large; random distribn\\
	(c)			&	$3$			&	$0.005$		&	random from single gene in $\mathcal{G}_{\phi}$
												&	$S$ small; $\delta_k$ large; single gene\\
	(d)			&	$10$			&	$0.001$		&	random from $\mathcal{G}_{\phi}$
												&	$S$ large; $\delta_k$ small; random distribn\\
	(e)			&	$3$			&	$0.001$		&	random from $\mathcal{G}_{\phi}$
												&	$S$ small; $\delta_k$ small; random distribn\\
	(f)			&	$3$			&	$0.001$		&	random from single gene in $\mathcal{G}_{\phi}$
												&	$S$ small; $\delta_k$ small; single gene\\
	\end{tabular}
	\caption{Scenarios tested in simulation study.  For scenarios (c) and (f), in the rare event that a gene has less than $3$ SNPs, all SNPs within the gene are selected.}
	\label{tab:simulationScenarios}
	\end{center}
\end{table}

The list of scenarios tested is presented in Table \ref{tab:simulationScenarios}.  First, we consider scenarios where the number of causal SNPs is small ($S = 3$) or large ($S = 10$).  Secondly, we consider two different SNP effect sizes.  We choose values for $\sigma^2_{\epsilon}$ and $\delta_k$ to mimic effect sizes obtained in recent association studies, focussing particularly on the smallest reported effect sizes.  \cite{Park2010} review GWAS for a number of quantitative traits (height, Crohn's disease and breast, prostate and colorectal cancers) and report values for $\text{ES}_k$  ranging from $0.02$ to $0.0004$.  \cite{Cho2009} report values for $\zeta_k$ for 8 quantitative traits in a large GWAS ranging from $1.6$ to $0.006$.  A recent neuroimaging genetic study measuring genetic effects on a variety of traits related to brain structure reports significant values for $\zeta_k$ of around $0.07$ \citep{Joyner2009}.  We set $\sigma_{\epsilon} = 0.2$, and test $\delta_k = 0.005$ and $0.001$, which gives values for $\text{ES}_k = 0.001$ and $0.00004$ and $\text{E}(\zeta_k) = 0.01$ and $0.002$ respectively.  Finally, we also vary the particular distribution of SNPs with respect to their location within causal pathways.  We expect the distribution of causal SNPs with respect to genes and associated LD blocks to affect performance, both in our regression model, and in the case where pathway scores are derived in a two-step process that begins with the calculation of gene association scores \citep{Wang2007}.  The distributions of $|\mathcal{C}|$, the number of causal pathways for each scenario, are shown in Fig.~\ref{fig:nCausalPaths}.  
% distributions of number of causal pathways
\begin{figure}[hbt]
	\begin{center}
	\includegraphics[trim = 40mm 10mm 40mm 20mm, clip, scale=0.33]{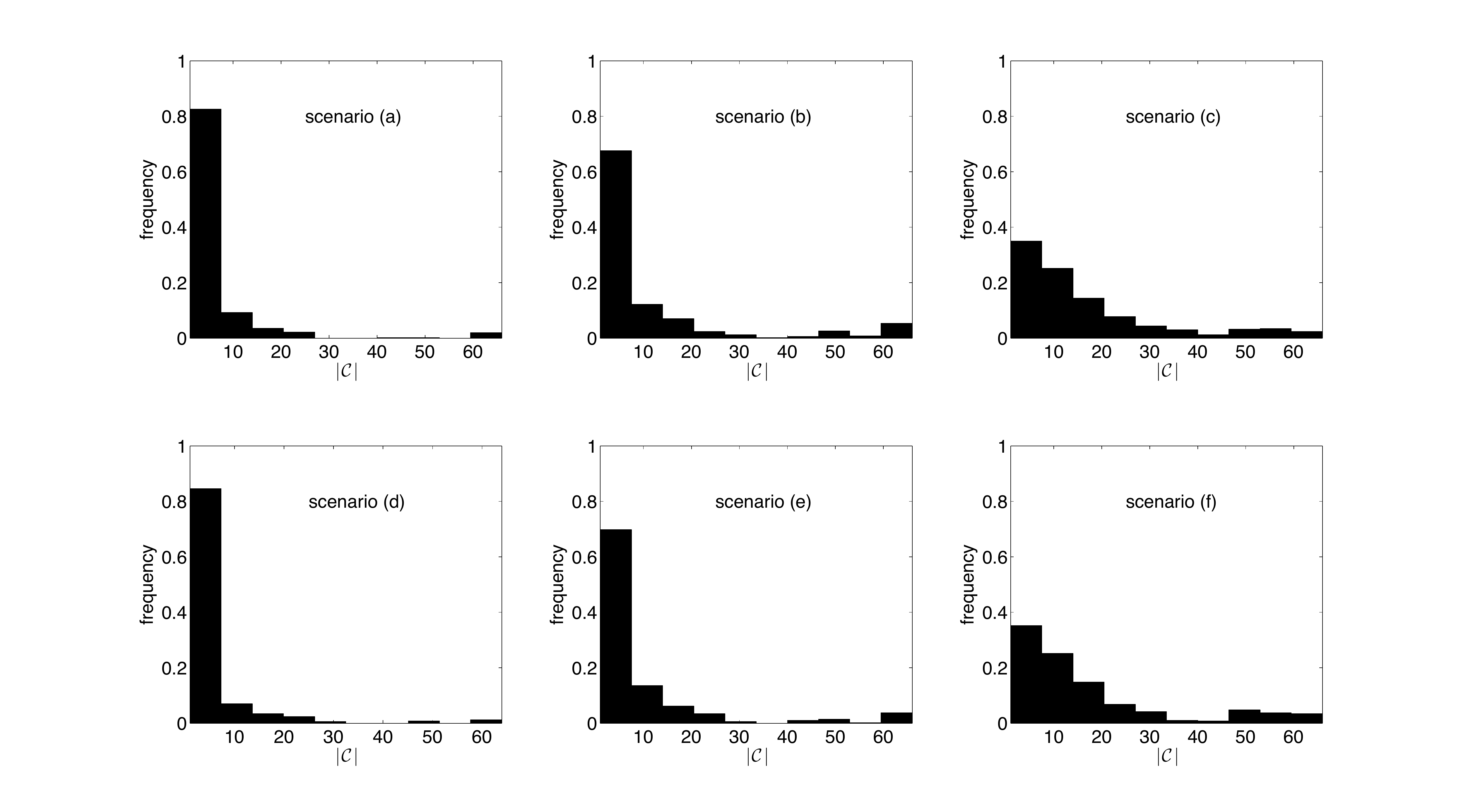}
	\caption{Distributions of $|\mathcal{C}|$ across $500$ MC simulations for the $6$ scenarios described in Table \ref{tab:simulationScenarios}.  Where SNPs are distributed within a single gene (scenarios (c) and (f)), the number of causal pathways tends to be larger, since a single gene can map to multiple pathways.  Where SNPs are distributed randomly across $\mathcal{G}_{\phi}$ (scenarios (a), (b), (d), and (e)), this number tends to be smaller, particularly where the number of causal SNPs is large (scenarios (a) and (d)).}
	\label{fig:nCausalPaths}
	\end{center}
\end{figure}

%%%%%%%%%%%%%%%%%%%%%%%%%%%%%%%%%%%%%%%%%%%%%%%%%%%%%%%%%%%%%%%%%%%%%%%%%%
\section{Results} \label{sec:Results}

% table of execution times
\begin{table}
	\begin{center}
	\begin{tabular}{l||cc|cc|cc}
	\hline
				& \multicolumn{2}{c|}{$P^*=4$k, $L=126$} & \multicolumn{2}{c|}{$P^*=66$k, $L=551$} & \multicolumn{2}{c}{$P^*=647$k, $L=879$}\\
	sample size	& BCD & BCD+ & BCD & BCD+ & BCD & BCD+\\
	\hline
	$371$ $(N/2)$	&	7.93	&	0.17	&	421	&	1.35	&	5490		&	16\\
	$743$ $(N)$	&	16.9	&	0.27	&	511	&	2.5	&	6430		&	30.0\\
	\hline
	\end{tabular}
	\caption{GL estimation times (seconds) with $M = 10$.  Table shows the time taken for the full estimation with a null $\mathcal{N} \sim (0,1)$ response, and with varying number of SNPs ($P^*$), and sample size ($N$).  `BCD' - estimation using block coordinate descent only.  `BCD+' - estimation using BCD with active set, Taylor approximation of the group penalty and efficient computation of block residuals.  Genotype and pathway data as described in section \ref{subsec:genotypePathwayData}.  
$P^*= 4k: 5,000$ SNPs from chromosome 1 mapped to $126$ pathways.
$P^*= 66k:$ all $33,850$ genotyped SNPs from chromosome 1 mapped to $551$ pathways.
$P^*= 647k: 448,294$ genome-wide SNPs mapped to $879$ pathways.  All computations performed using multi-threading on a single machine with $8$ $3.2$ GHz processors and $64$GB RAM.}
	\label{tab:executionTimes}
	\end{center}
\end{table}

We begin with an investigation of the effect of our proposed speed ups to the GL estimation algorithm.  We first note that GL estimation times will depend on the sample size $(N)$ and the number of SNPs $(P)$, which will in turn affect the number of mapped pathways $(L)$ and $P^*$.  Estimation times will further depend on the number of groups selected $(M)$, and the amount of signal present, since this affects convergence times.  For illustrative purposes, in Table \ref{tab:executionTimes} we show gains in execution time compared with `standard' block coordinate descent, using our proposed speed ups for a single model fit with a null response, and for $M = 10$.  Estimation times are seen to be substantially reduced across a range of values for $N$ and $P$, dramatically so for larger datasets.

% illustration of bias-adjusted adaptive weighing procedure
\begin{figure}
	\begin{center}
	\subfigure[]{\includegraphics[scale=0.26]{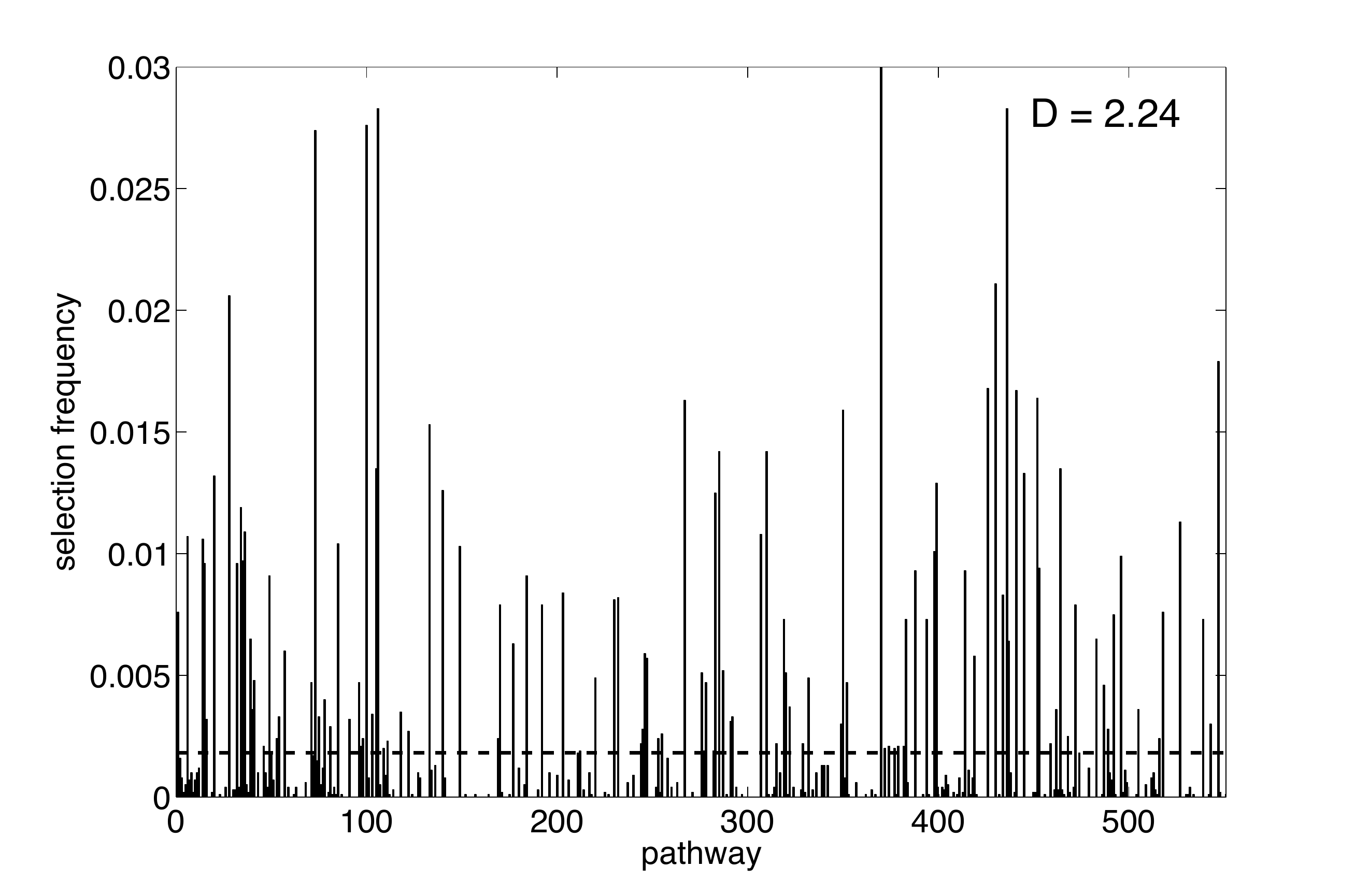}}	
	\subfigure[]{\includegraphics[scale=0.26]{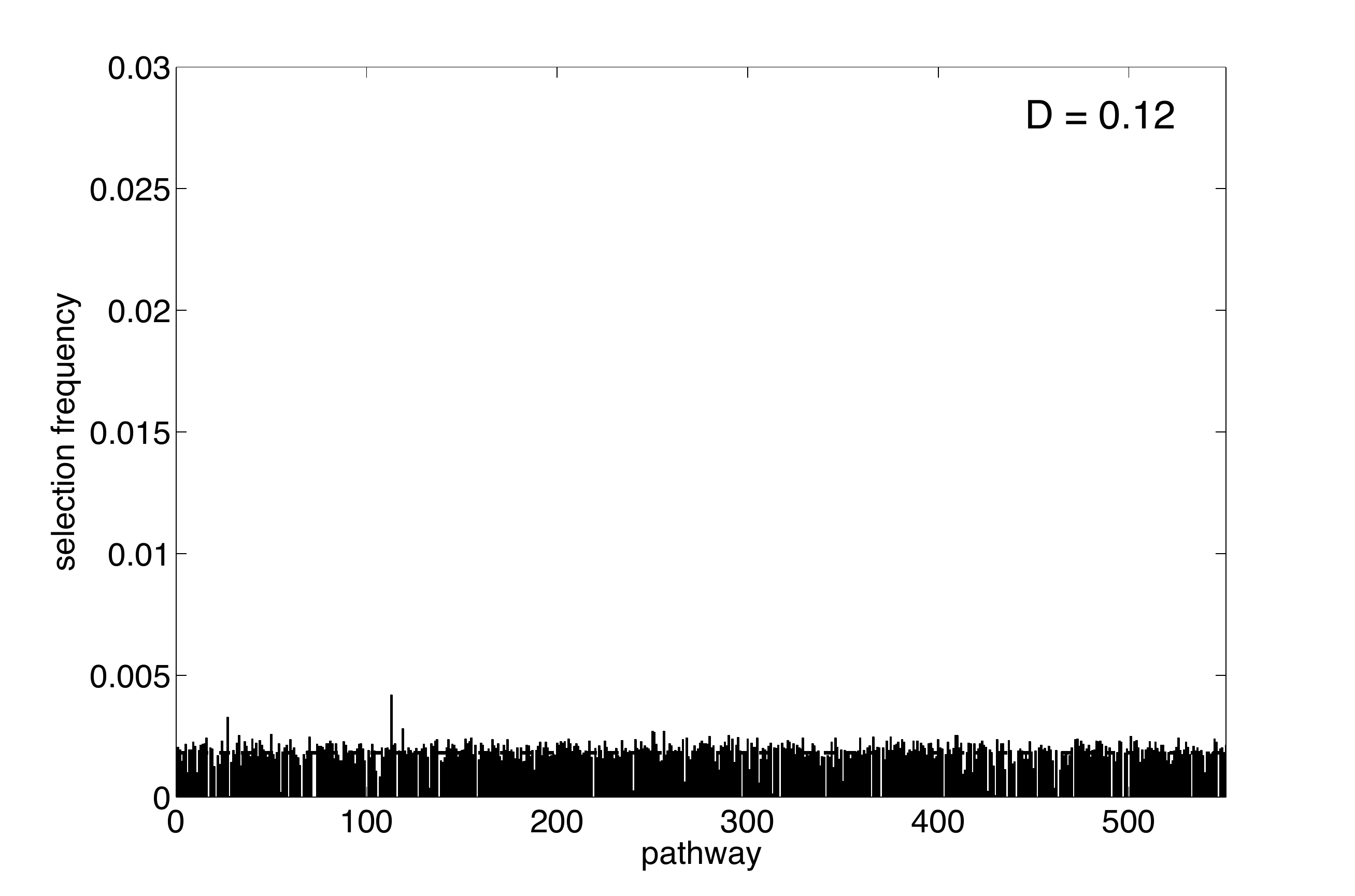}}
	\subfigure[]{\includegraphics[scale = 0.26]{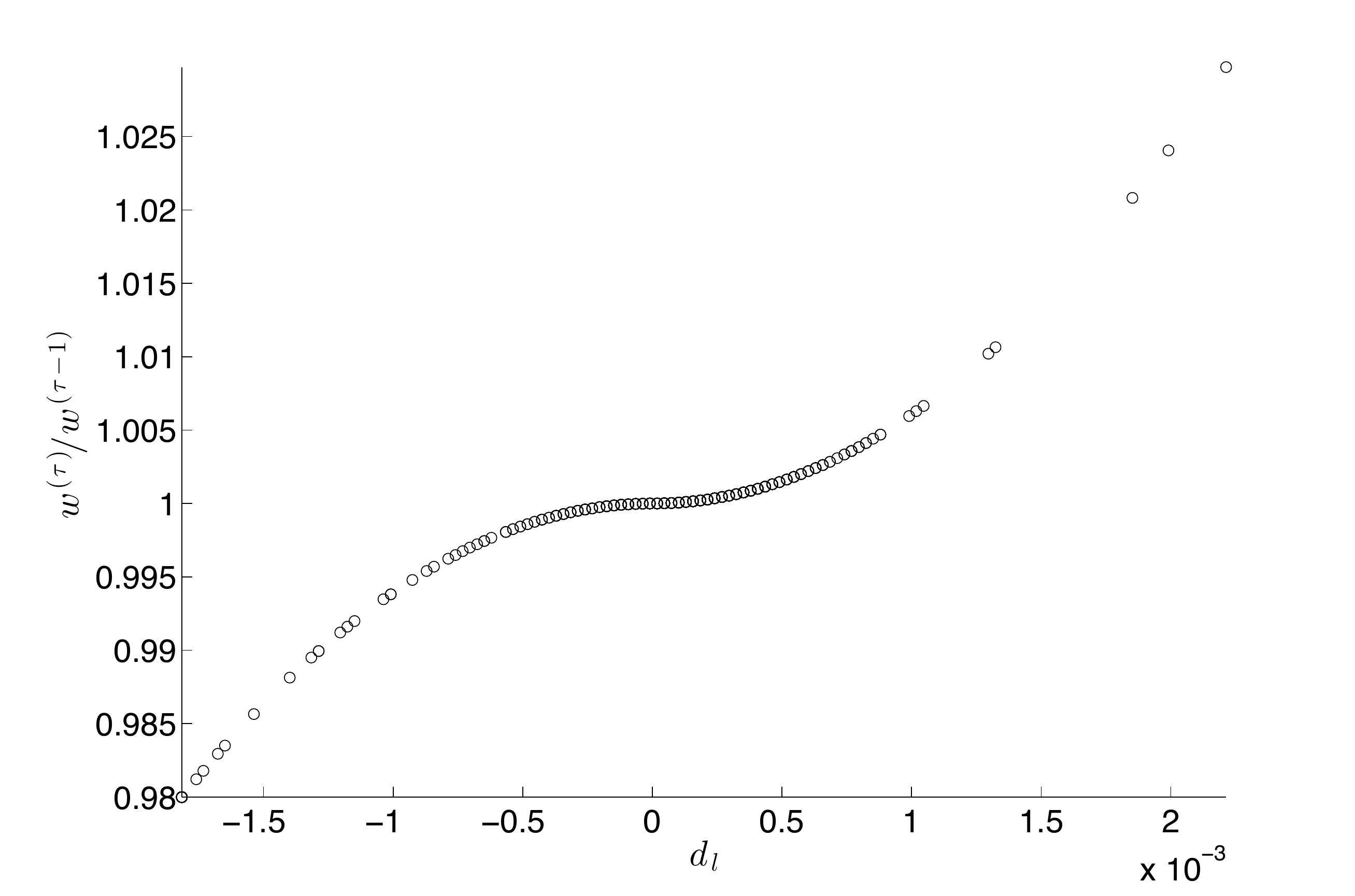}}	
	\subfigure[]{\includegraphics[scale = 0.26]{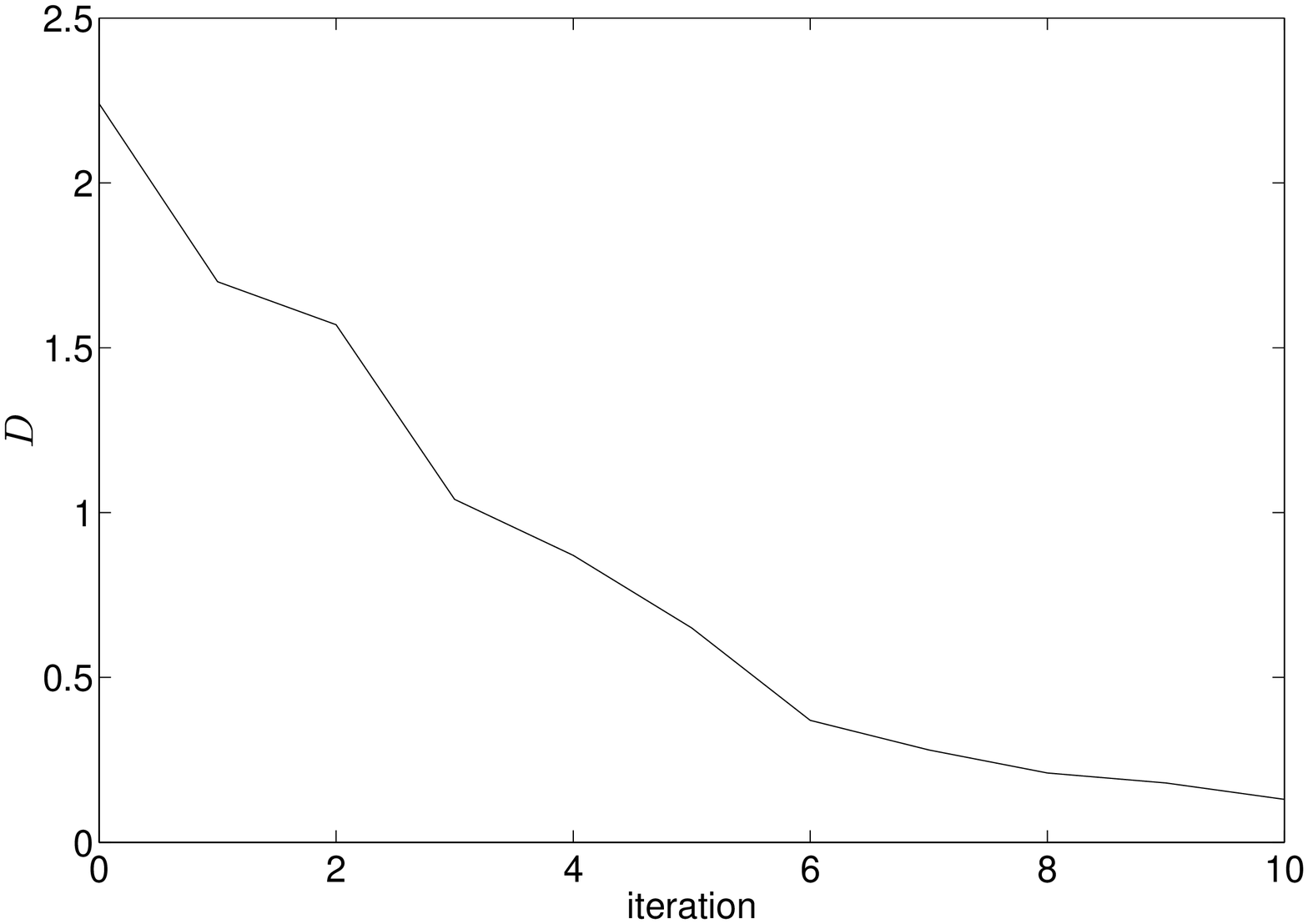}}
	\caption{Application of bias-adjusted weighting procedure to the data used in the simulation study.  $R = 40,000$, with a different null response,  $y \sim \mathcal{N}(0,1)$, at each MC simulation.  $\alpha = 0.98$. (a) Empirical pathway selection frequency distribution, $\Pi^*$, with standard, pathway size weighting, $w_l = \sqrt{S_l}$.  $D = 2.24$.  Dotted horizontal line shows the expected distribution, $\Pi_l = 1/L \simeq 0.002$.  (b) $\Pi^*$ with bias-adjusted weights after 10 iterations.  $D = 0.12$.  (c) Variation of weighting adjustment factor $\bm{w}^{(\tau)}/\bm{w}^{(\tau-1)}$ with $d_l$ at a single iteration, with $\alpha=0.98$.  Each point represents the adjustment to a single $w_l, l = 1,\ldots,L$. (d) Decrease in K-L divergence, $D$, over 10 iterations.}
	\label{fig:adaptiveWeights}
	\end{center}
\end{figure}

We next turn to the application of P-GLAW to real genotype and pathway data described in section \ref{subsec:pathwayWeighting}.  We apply this procedure over 10 iterations, each with $R=40,000$ MC simulations with a response $y \sim \mathcal{N}(0,1)$.  Fig.~\ref{fig:adaptiveWeights} (c) shows how the weight adjustment factor $\bm{w}^{(\tau)}/\bm{w}^{(\tau-1)}$, (see \eqref{eq:adjustWeights}), varies with $d_l$ across all pathways at a single iteration.  Fig.~\ref{fig:adaptiveWeights} (a) and (b) shows the observed, empirical distribution, $\Pi^*$, using the standard size weighting (\ref{eq:std_size_weighting}), and the adapted weights (\ref{eq:adjustWeights}) after 10 iterations, respectively.  The corresponding KL divergence measure, $D$, is observed to reduce steadily over the 10 iterations (Fig.~\ref{fig:adaptiveWeights} (d)), illustrating how the proposed weight adjustment procedure reduces pathway selection bias.\\

For the remainder of this section, we assess the performance of our proposed P-GLAW method using simulated phenotypes under the simulation framework described in the previous section, and using the bias-adjusted pathway weights described above.  We first compare performance using the bias-adjusted weights with that obtained using the standard size weighting \eqref{eq:std_size_weighting}.  We find the adjusted weighting scheme offers a considerable improvement in ranking performance for all ranking measures, and illustrate this in Fig.~\ref{fig:compareWeightingSchemes} for a single scenario (scenario (a)) using the ranking performance measures described in section \ref{subsec:PathwayRanking}.  Fig.~\ref{fig:compareWeightingSchemes} (a) shows the first ranking measure ($r_{k_1}$) as a ROC curve, in which we show the proportion of simulations with $r_{k_1} \le z$, for ranks $z = 1, 2, \ldots, 100$.  We plot $z$ on the horizontal axis as a false positive rate (FPR), so that $\mbox{FPR} = (z-1)/L$.  At a FPR of $0.05$, we see that the adapted weighting scheme shows a more than 2 fold increase in power (from $0.29$ to  $0.62$) over the standard pathway size weighting \eqref{eq:std_size_weighting}, indicating $62\%$ of MC simulations have $r_{k_1} \le 28$, compared with $29\%$ for the standard size weighting.  The distribution of $p_{100}$ across $500$ MC simulations is illustrated as a boxplot in Fig.~\ref{fig:compareWeightingSchemes} (b). Here we see that the adapted weighting scheme offers a clear and substantial improvement in GL's capacity to rank a high proportion of causal pathways in the top $100$ ($p = 2.03 \times 10^{-50}$ that the two population $p_{100}$ CDFs are equal using a two-sample Kolmogorov-Smirnov (KS) test).  GL with the standard weighting scheme performs particularly poorly with $55\%$ of simulations failing to rank any causal pathway in any simulation, compared with $18\%$ for the adapted weighting scheme.  Finally, Fig.~\ref{fig:compareWeightingSchemes} (c) shows the distribution of the $R$ ranking measure across $500$ simulations under the two weighting schemes.  Once again we see that the adaptive weighting scheme demonstrates improved ranking performance over the standard size weighting scheme, with the distribution of $R$ scores skewed towards lower values for the former, indicating that causal pathways tend to be ranked higher.

% compare adapted weights to standard size weighting
\begin{figure}
	\begin{center}
	\subfigure[]{\includegraphics[trim = 10mm 60mm 10mm 60mm, clip, scale=0.33]{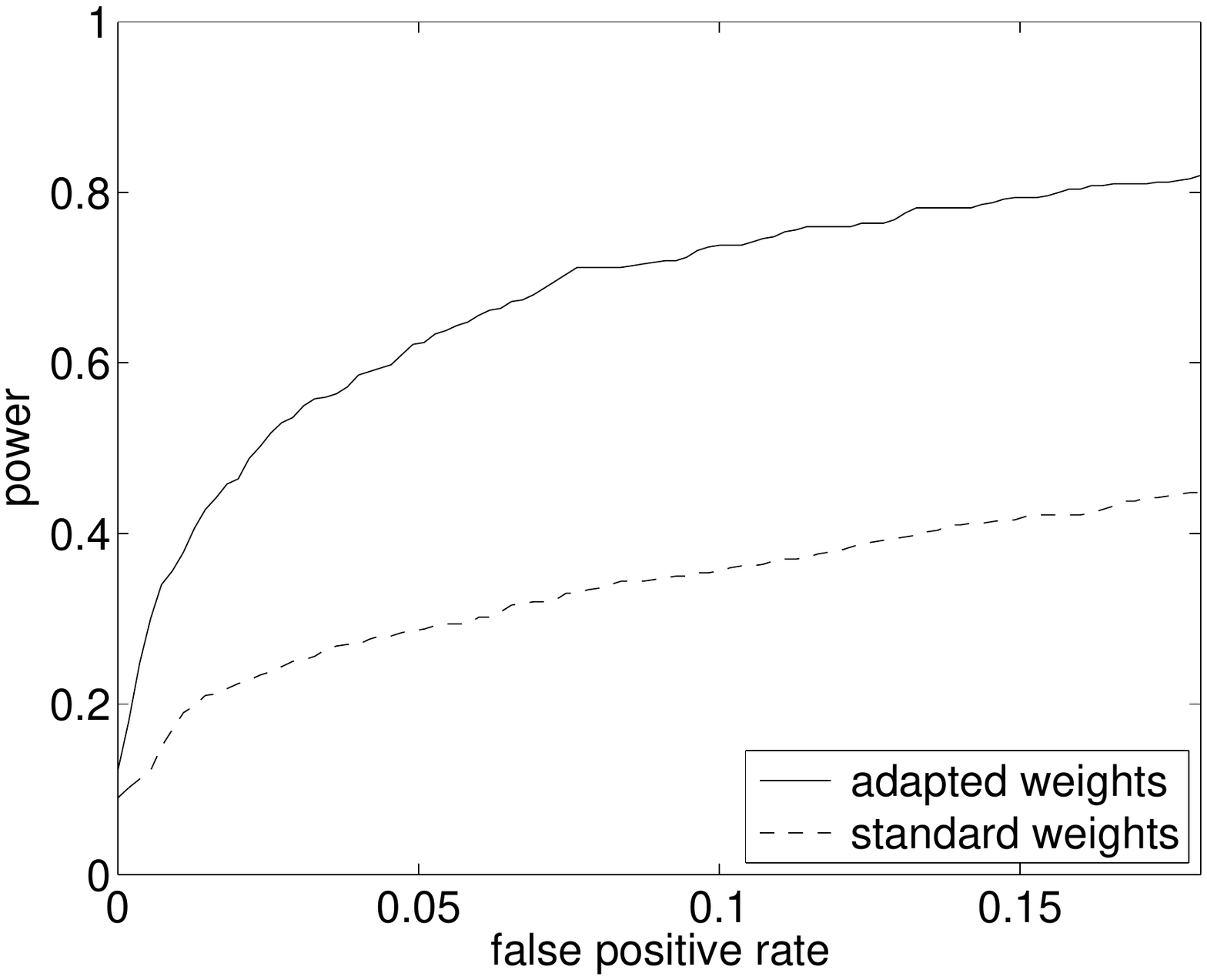}}
	\subfigure[]{\includegraphics[trim = 10mm 60mm 10mm 60mm, clip, scale=0.33]{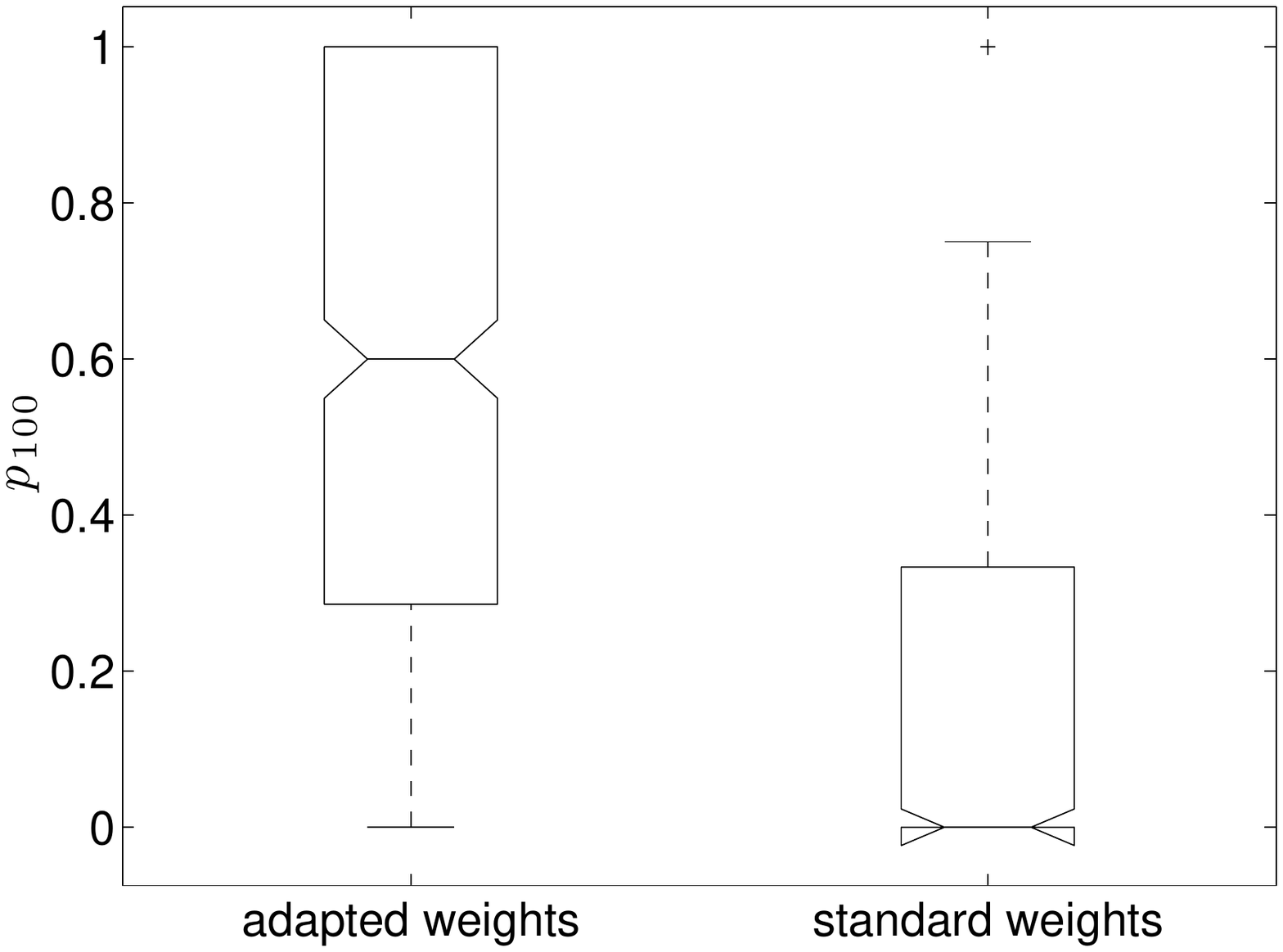}}
	\subfigure[]{\includegraphics[trim = 10mm 60mm 10mm 60mm, clip, scale=0.33]{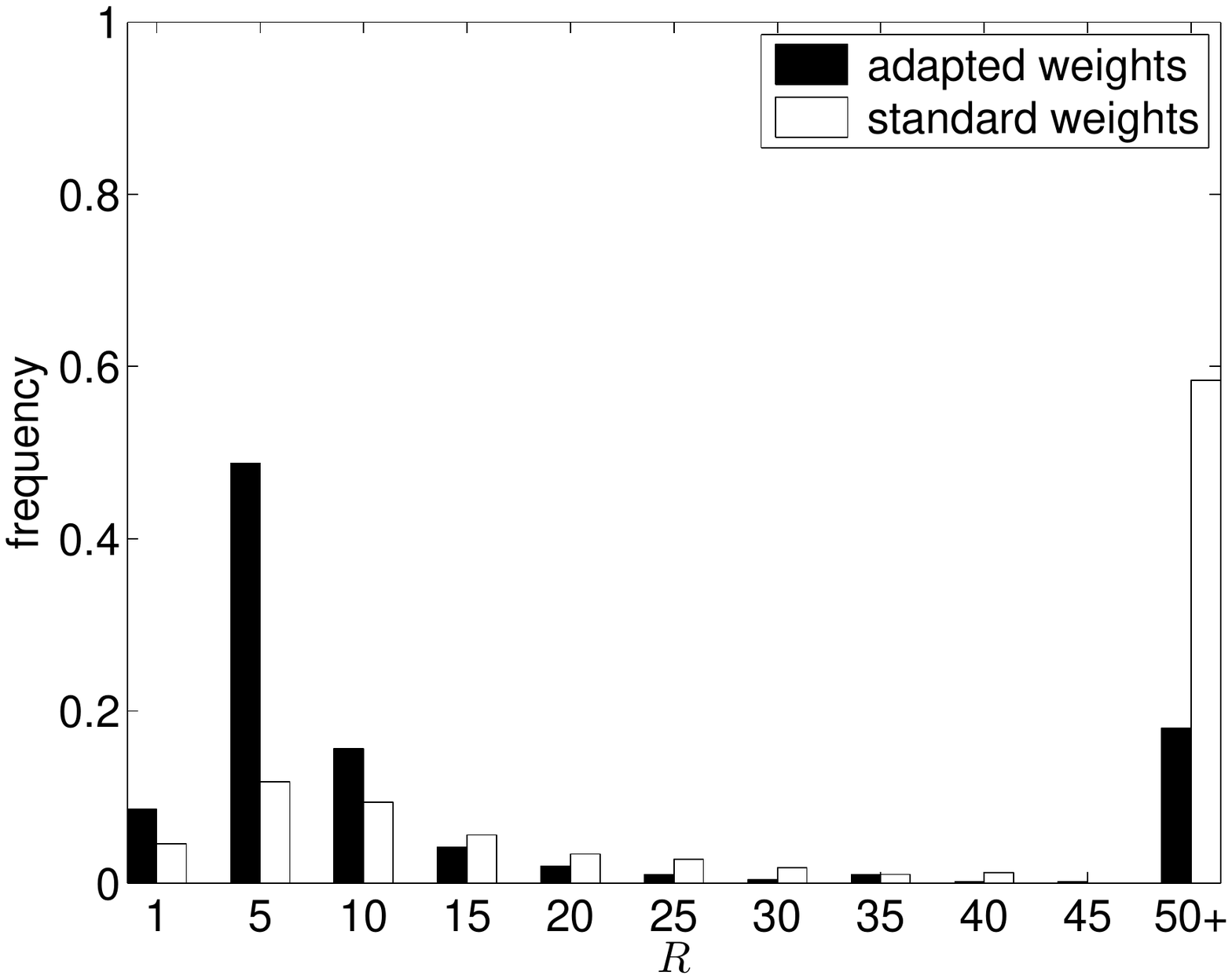}}		
	\caption{Comparison of ranking performance: adaptive weighting scheme (section \ref{subsec:pathwayWeighting}) vs.~standard pathway size weighting \eqref{eq:std_size_weighting}.  $S = 10; \delta_k = 0.005$; SNPs randomly distributed across $\mathcal{G}_{\phi}$.  (a) ROC curves illustrating power to identify at least one causal pathway in the top $100$.  Power is average across $500$ simulations. (b) Distribution of ranking power, $p_{100}$, across $500$ simulations.  This is the proportion $|\mathcal{C}|^*_{100}| / |\mathcal{C}|$ of causal pathways in $\mathcal{C}$ that are ranked in the top $100$ pathways.  Notches indicate $95\%$ confidence intervals for the true median. (c) Distribution of the power-adjusted, normalised, weighted ranking score, $R$, across $500$ simulations.  The final `$50+$' column includes simulations for which no causal pathway was ranked in the top $100$, i.e.~$\mathcal{C}^*_{100} = \emptyset; R = 100$.}
	\label{fig:compareWeightingSchemes}
	\end{center}
\end{figure}

We next assess P-GLAW ranking performance with the adapted weighting scheme across the full range of scenarios, and compare these with pathway rankings obtained using the method proposed by \citet{Wang2007}, commonly referred to as `GenGen' (GG).  GG is a widely-used, GSEA-type PGAS method that measures pathway enrichment using genes scores derived from univariate SNP statistics.  Studies using GG include searches for implicated pathways in Crohn's disease \citep{Wang2009}, autism spectrum disorders \citep{Wang2009a}, breast cancer \citep{Menashe2010} and Alzheimer's disease \citep{Lambert2010}.  GG begins by scoring each SNP according to its association with the phenotype.  SNPs are then mapped to genes within a specified distance, and each gene is scored according to its most significant mapped SNP.  The enrichment of highly-ranked genes in a given pathway is then compared with those in all other pathways, to obtain a pathway enrichment score. For GenGen we use identical source data (genotypes, phenotypes, SNP to gene, and gene to pathway mappings), and rank pathways by normalised enrichment score, determined from 1,000 permutations (the GG default settings).  MC simulations for P-GLAW and GG are performed in parallel across 50 (P-GLAW) and 500 (GG) processors respectively, on a high-performance computing cluster.  As described above for alternative weighting schemes, results for the comparison study are presented in the form of $r_{k_1}$ ROC curves (Fig.~\ref{fig:ROCs}), $p_{100}$ boxplots (Fig.~\ref{fig:p100}) and $R$ bar graphs (Fig.~\ref{fig:R100}).  Selected ranking measures are presented in numerical form in Tables \ref{tab:Results1} and \ref{tab:Results2}.\\

% first results table
\begin{table}[htbp]
\begin{center}
	\scriptsize
	\tabcolsep 4pt
\begin{tabular}{@{}c||ccc|ccc|ccc|c@{}}
	\hline
	scen. & \multicolumn{3}{c|}{ROC power, fpr = $0.05$} & \multicolumn{3}{c|}{median $p_{100}$} & \multicolumn{3}{c|}{propn. $p_{100}=0$} & KS 2 sample test \\
	 & P-GLAW & GG & ratio & P-GLAW & GG & ratio & P-GLAW & GG & ratio & $p_{100}$ cdfs the same\\
	 \hline
	(a) & 0.62 & 0.35 & 1.76 & 0.60 & 0.60 & 1.00 & 0.18 & 0.26 & 0.70 & p = 0.0082 \\
	(b) & 0.61 & 0.33 & 1.84 & 0.33 & 0.11 & 3.00 & 0.21 & 0.45 & 0.46 & p = $9.6\times 10^{-25}$ \\
	(c) & 0.81 & 0.54 & 1.49 & 0.35 & 0.20 & 1.73 & 0.06 & 0.23 & 0.25 & p = $2.5\times 10^{-25}$ \\
	(d) & 0.44 & 0.18 & 2.37 & 0.33 & 0.00 & $\infty$ & 0.30 & 0.62 & 0.48 & p = $7.7\times 10^{-27}$ \\
	(e) & 0.59 & 0.27 & 2.18 & 0.33 & 0.01 & 37.33 & 0.23 & 0.50 & 0.46 & p = $9.2\times 10^{-28}$ \\
	(f) & 0.79 & 0.45 & 1.74 & 0.31 & 0.14 & 2.31 & 0.06 & 0.31 & 0.20 & p = $3\times 10^{-38}$ \\
\end{tabular}
\caption{Selected ranking performance measures for P-GLAW and GG for the 6 scenarios described in Table \ref{tab:simulationScenarios}.  \emph{ROC power, fpr = $0.05$}: proportion of $500$ MC simulations with $r_{k_1} \le 28$ corresponding to a fpr of $0.05$.  \emph{median $p_{100}$}: median of $p_{100}$ distribution across $500$ MC simulations.  \emph{Proportion with $p_{100}=0$}: proportion of $500$ MC simulations with no causal pathway in the top $100$ ranks.  \emph{KS 2 sample test:} two-sample Kolmogorov-Smirnov test of the hypothesis that the P-GLAW and GG $p_{100}$ population cdfs are the same.}
\label{tab:Results1}
\end{center}
\end{table}

Beginning with the ROC curves illustrating the $r_{k_1}$ ranking measure (Fig.~\ref{fig:ROCs} and first 3 columns of Table~\ref{tab:Results1}), GG consistently demonstrates increased power and specificity across all of the top $100$ ranks illustrated.  In addition, the relative gain in power for P-GLAW is greater at the smallest effect size for each equivalent scenario, (a) vs.~(d), (b) vs.~(e), and (c) vs.~(f).  At the smaller effect size, where causal SNPs are distributed randomly within causal pathways, power increases where the number of causal SNPs is fewer ((d) vs.~(e)).  Finally, maximum power is achieved for both methods where causal SNPs are located within a single gene ((c) and (f)).

Turning to the distributions of the $p_{100}$ ranking measure (Fig.~\ref{fig:p100}, and columns 4 to 9 in Table~\ref{tab:Results1}), P-GLAW again outperforms GG across all scenarios.  For example, the null hypothesis that the two population cdfs are equal is rejected at the $\alpha = 0.05$ level (Table \ref{tab:Results1}, final column), as is the null hypothesis that the two sample medians are the same (Fig.~\ref{fig:p100}), except for scenario (a) where median $p_{100}$ is not significantly different for the two methods.  Excluding scenario (a) where both methods perform relatively well, P-GLAW median $p_{100}$ is consistent across each scenario, and is maintained from the larger to the smaller effect size.  This is in marked contrast to GG, where this measure shows a large decrease at the smaller effect size, although the decrease is less marked when causal SNPs are located within a single gene.  A similar pattern persists for both P-GLAW and GG if we consider the proportion of simulations with $p_{100}=0$, i.e.~where no causal pathways are found in the top 100 ranks, except for P-GLAW in the case where causal SNPs are located in a single gene, where this measure is particularly low.

The final series of plots (Fig.~\ref{fig:R100}), illustrate the distributions of $R$ across all scenarios.  These distributions once again follow the trends in ranking performance highlighted above, but they offer a more nuanced view, in the sense that while this measure takes power into account, it is also sensitive to the actual causal pathway rankings.  Here we see that P-GLAW tends to rank causal pathways higher than GG, since all P-GLAW distributions are skewed towards lower $R$ values, indicating that causal pathways tend to be ranked higher.  This is borne out if we focus on the proportion of simulations with $R < 10$ (Table \ref{tab:Results2}, first 3 columns), which also illustrates how proportionate gains in ranking performance for P-GLAW over GG are largest for the smallest effect size ((a)-(c) vs.~(d)-(f)).  This table also gives results for the proportion of simulations demonstrating near optimal ranking of causal pathways ($R < 3$), although the very small frequencies suggest that little can be inferred from these.  

% second results table
\begin{table}[htbp]\footnotesize
\begin{center}
\begin{tabular}{c||ccc|ccc}
	\hline
	scenario & \multicolumn{3}{c|}{$R < 10$} & \multicolumn{3}{c}{$R < 3$} \\
	& P-GLAW & GG & ratio & P-GLAW & GG & ratio \\
	\hline
	(a) & 0.68 & 0.46 & 1.47 & 0.13 & 0.09 & 1.38 \\
	(b) & 0.50 & 0.24 & 2.11 & 0.03 & 0.03 & 0.93 \\
	(c) & 0.55 & 0.33 & 1.68 & 0.01 & 0.07 & 0.18 \\
	(d) & 0.44 & 0.20 & 2.22 & 0.03 & 0.02 & 2.00 \\
	(e) & 0.46 & 0.20 & 2.33 & 0.02 & 0.03 & 0.69 \\
	(f) & 0.45 & 0.23 & 1.96 & 0.01 & 0.04 & 0.30 \\
\end{tabular}
\end{center}
\caption{Proportion of $500$ simulations with $R < 10$ and $R < 3$ for the 6 scenarios described in Table \ref{tab:simulationScenarios}.}
\label{tab:Results2}
\end{table}

\renewcommand*{\thesubfigure}{} % gets rid of (a), (b) etc in subfigure captions as we want to change the order of these!

% ROCs
\begin{figure}[p]
	\begin{center}
	\subfigure[\scriptsize(a) $S = 10; \delta_k = 0.005;$ random distbn]{\includegraphics[trim = 10mm 60mm 10mm 60mm, clip, scale=0.30]{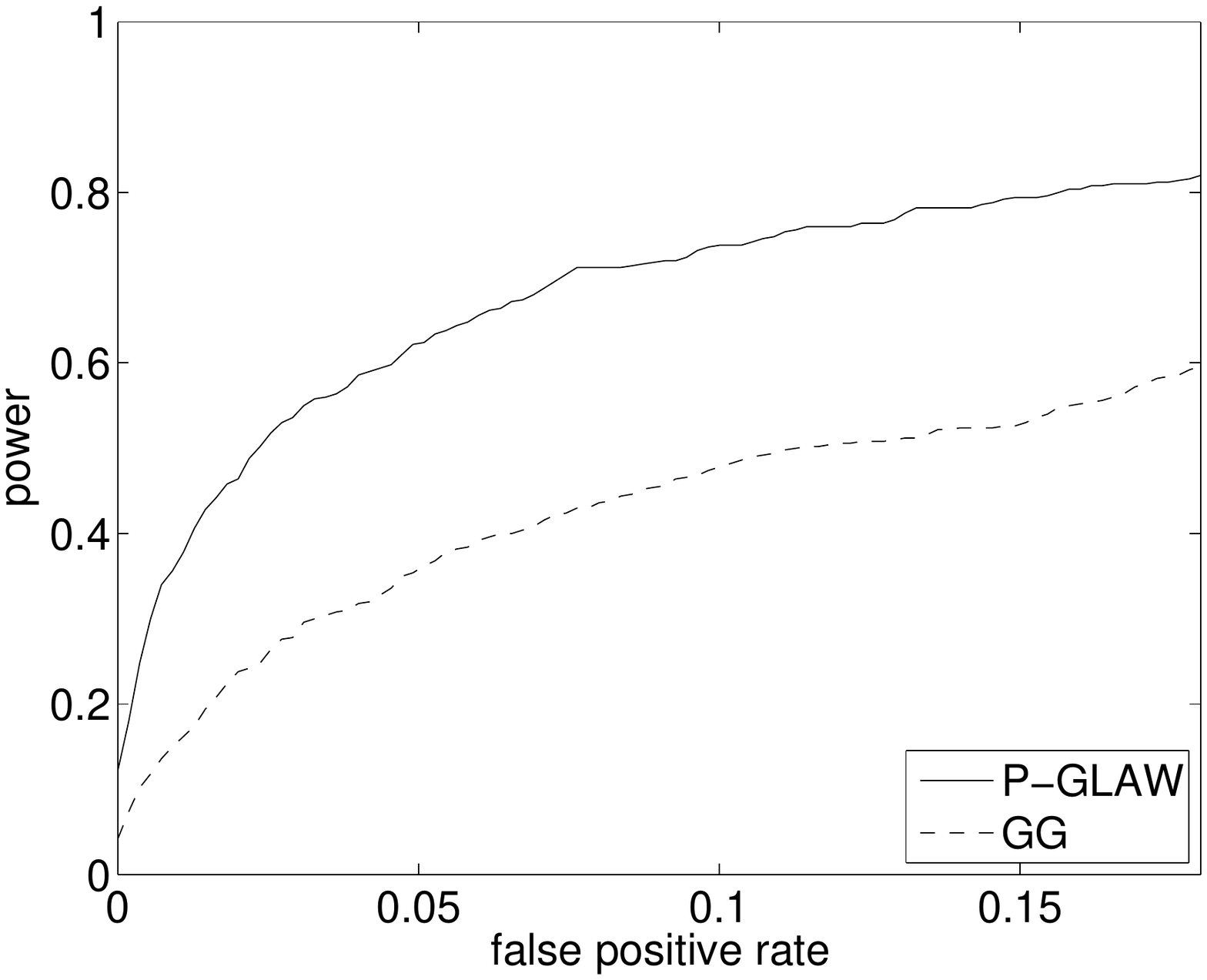}}
	\subfigure[\scriptsize(d) $S = 10; \delta_k = 0.001;$ random distbn]{\includegraphics[trim = 10mm 60mm 10mm 60mm, clip, scale=0.30]{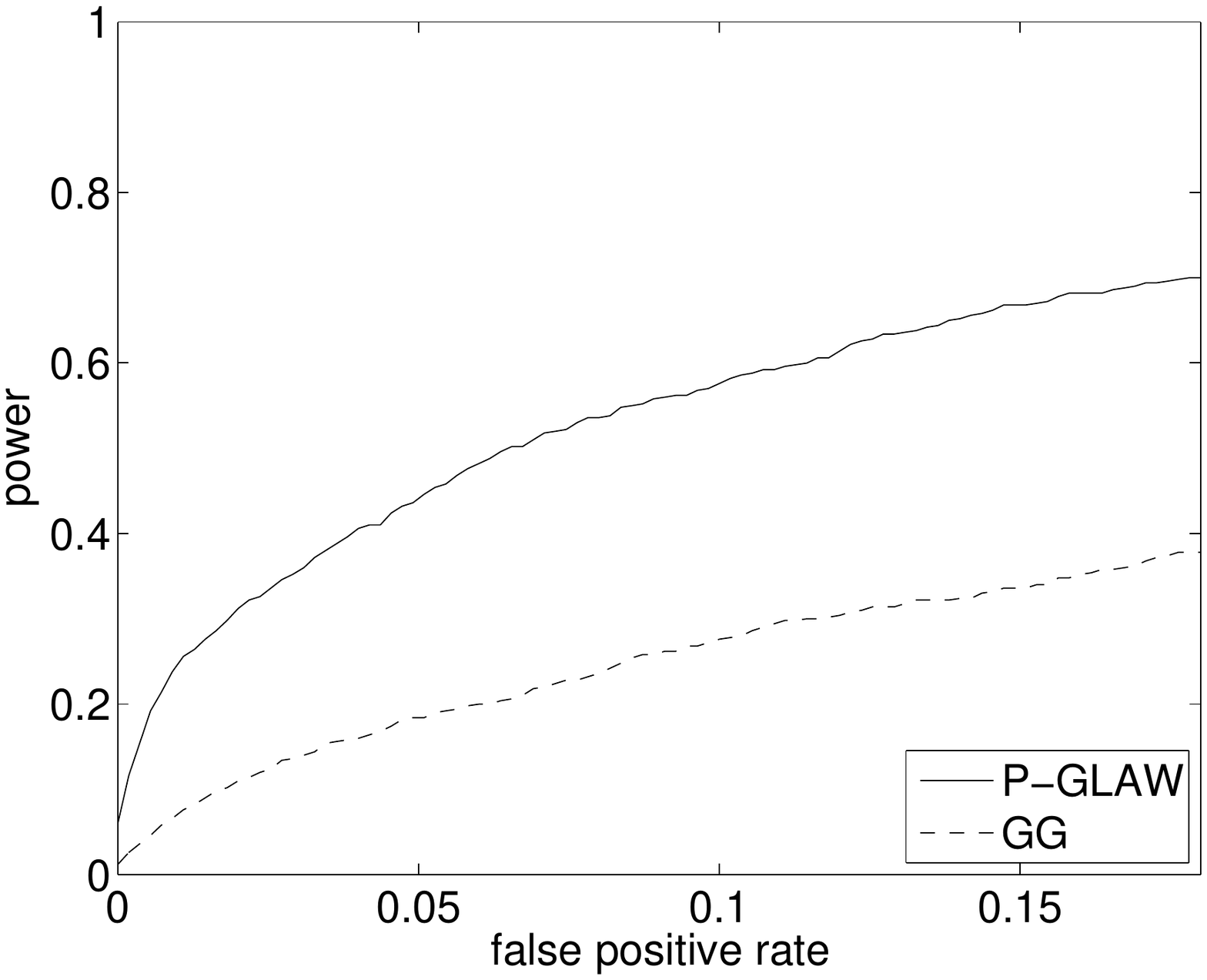}}
	\subfigure[\scriptsize(b) $S = 3; \delta_k = 0.005;$ random distbn]{\includegraphics[trim = 10mm 60mm 10mm 60mm, clip, scale=0.30]{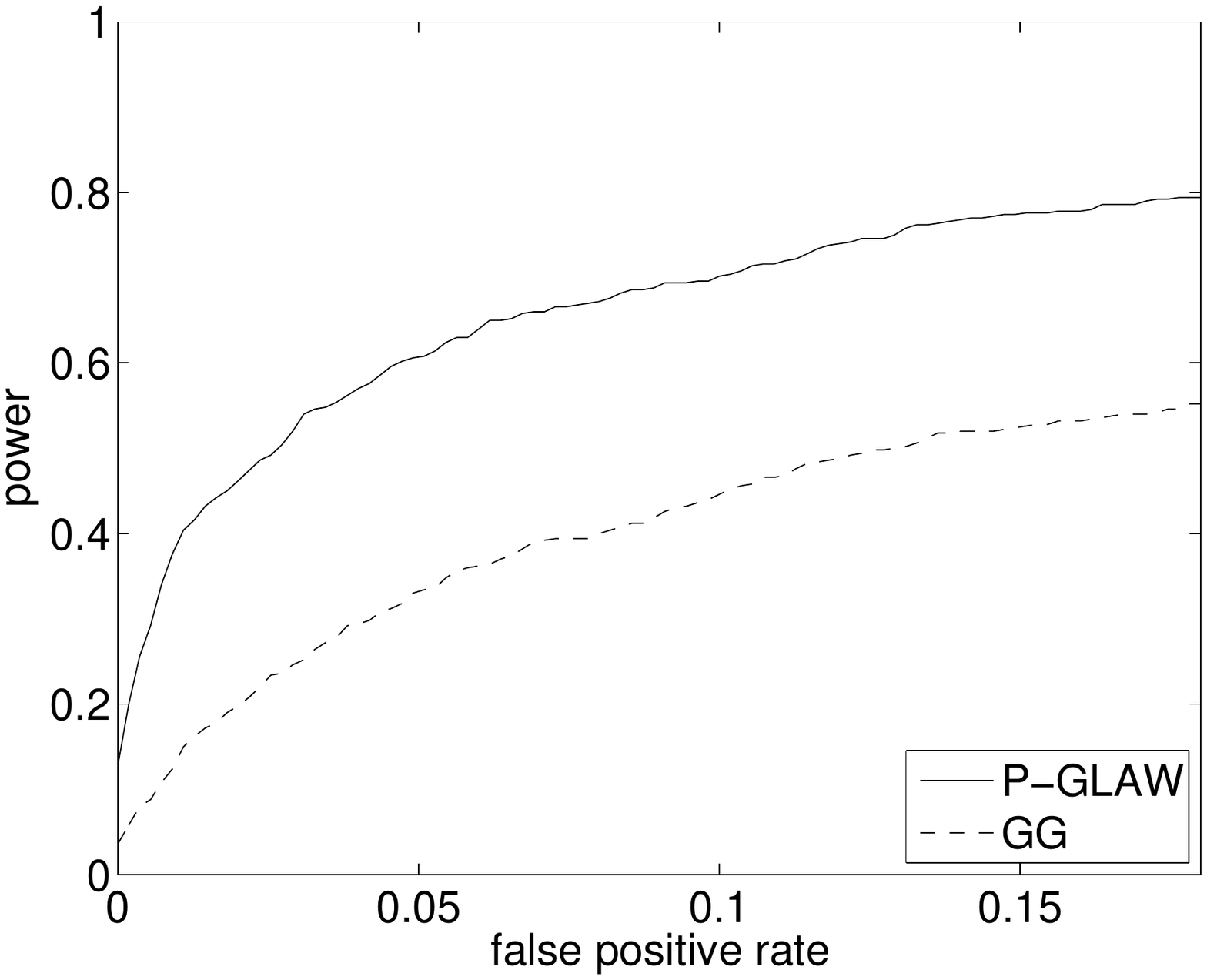}}
	\subfigure[\scriptsize(e) $S = 3; \delta_k = 0.001;$ random distbn]{\includegraphics[trim = 10mm 60mm 10mm 60mm, clip, scale=0.30]{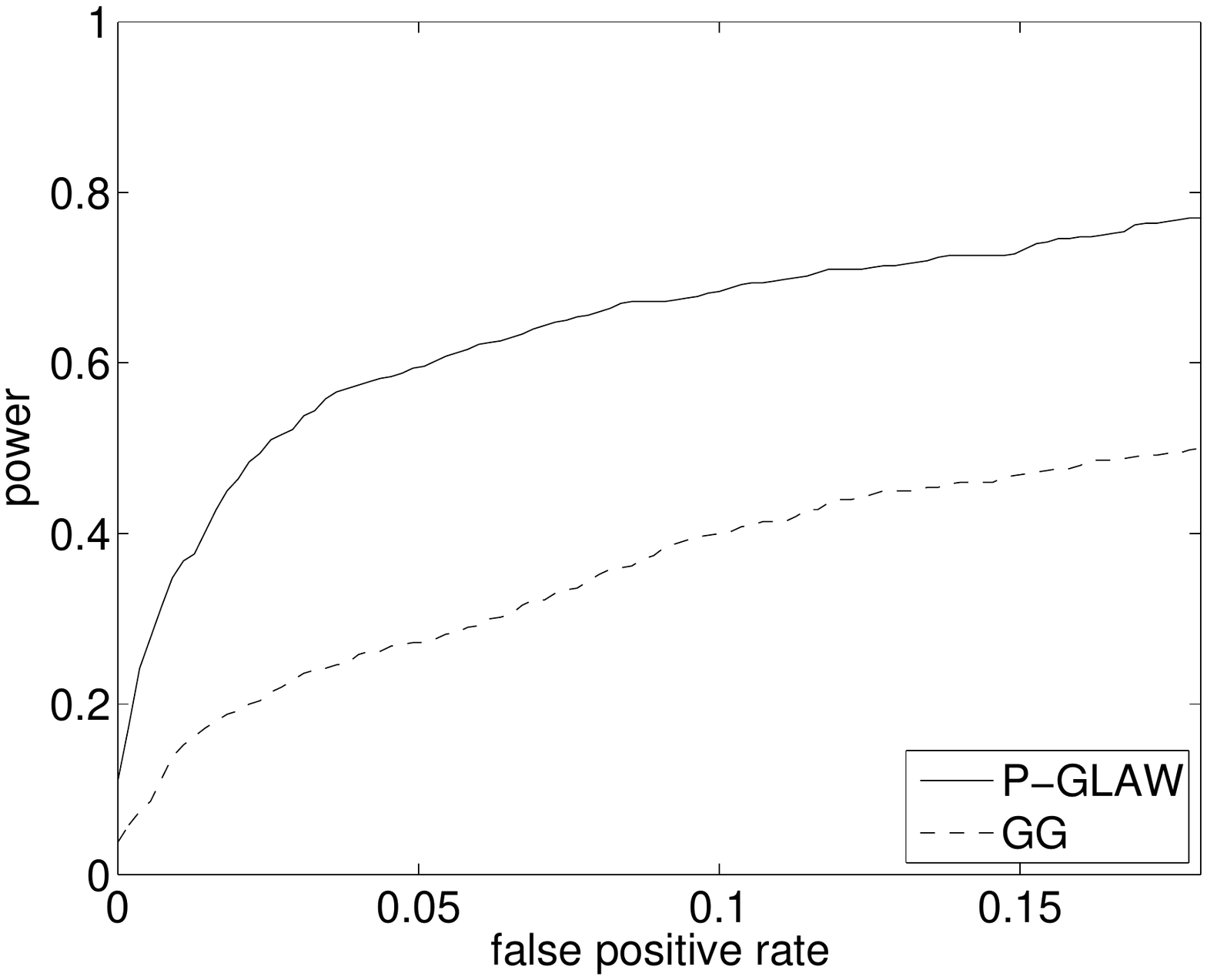}}
	\subfigure[\scriptsize(c) $S = 3; \delta_k = 0.005;$ single gene distbn]{\includegraphics[trim = 10mm 60mm 10mm 60mm, clip, scale=0.30]{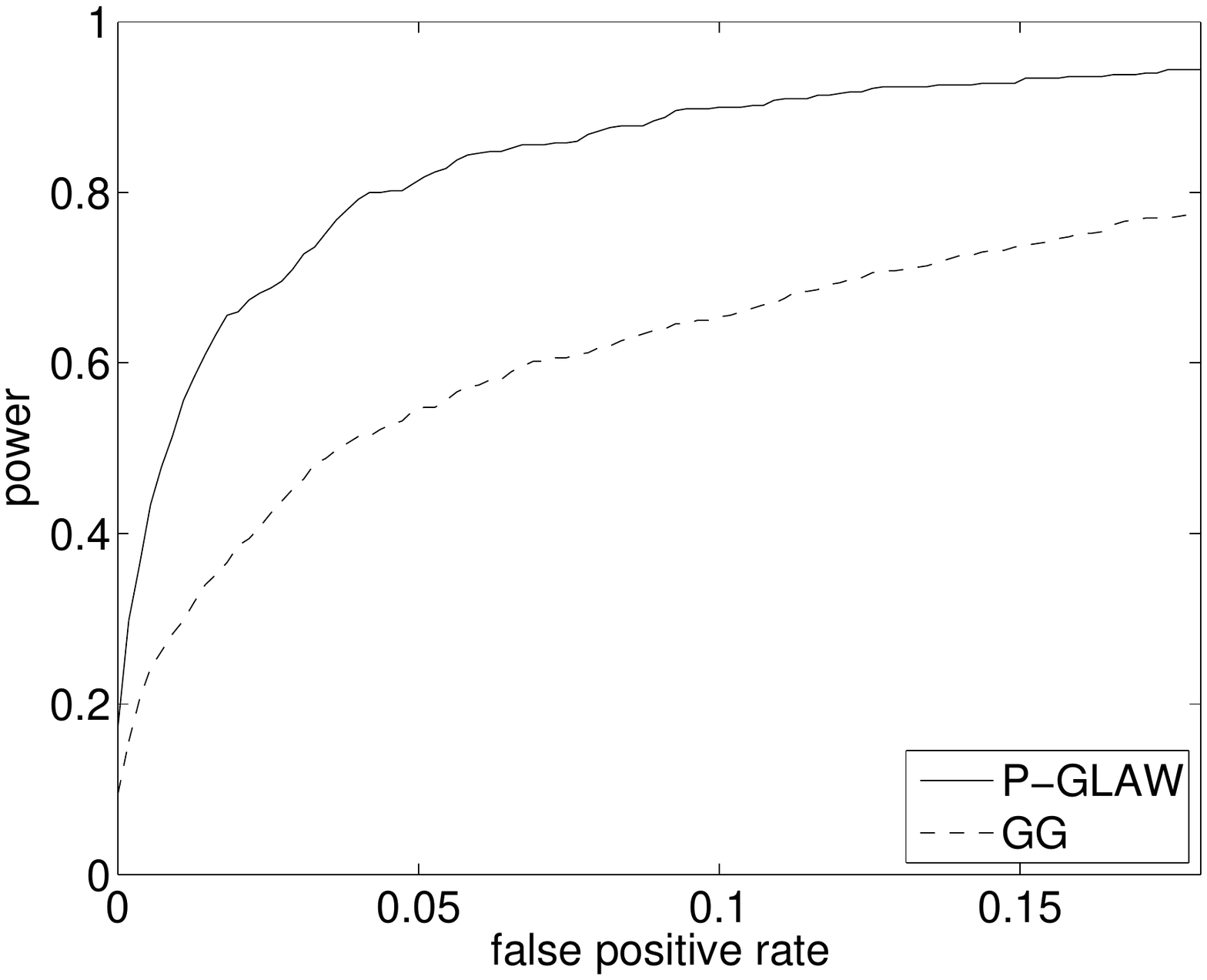}}
	\subfigure[\scriptsize(f) $S = 3; \delta_k = 0.001;$ single gene distbn]{\includegraphics[trim = 10mm 60mm 10mm 60mm, clip, scale=0.30]{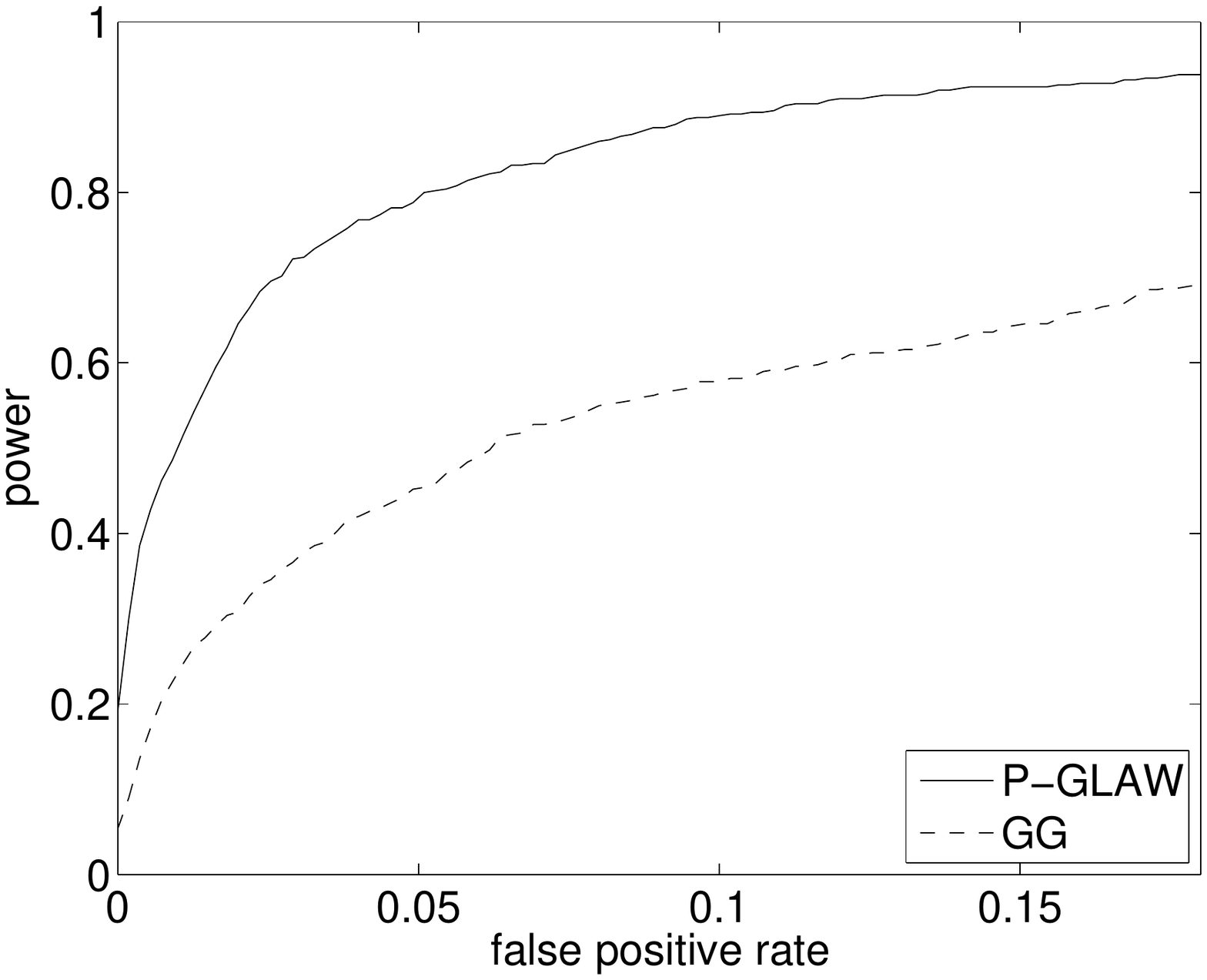}}	
	\caption{ROC curves illustrating proportion of simulations with $r_{k_1} \le z$, for ranks $z = 1, 2, \ldots, 100$.  Power is average across $500$ simulations.  $\mbox{False positive rate} = (z-1)/L$.  Scenarios corresponding to the higher SNP effect size ($\delta_k = 0.005$) are presented in the left-hand column, with the equivalent scenarios at the lower effect size ($\delta_k = 0.001$) on the right.}
	\label{fig:ROCs}
	\end{center}
\end{figure}

% power100 boxplots
\begin{figure}
	\begin{center}
	\subfigure[(a) $S = 10; \delta_k = 0.005;$ random distbn]{\includegraphics[trim = 10mm 60mm 10mm 60mm, clip, scale=0.32]{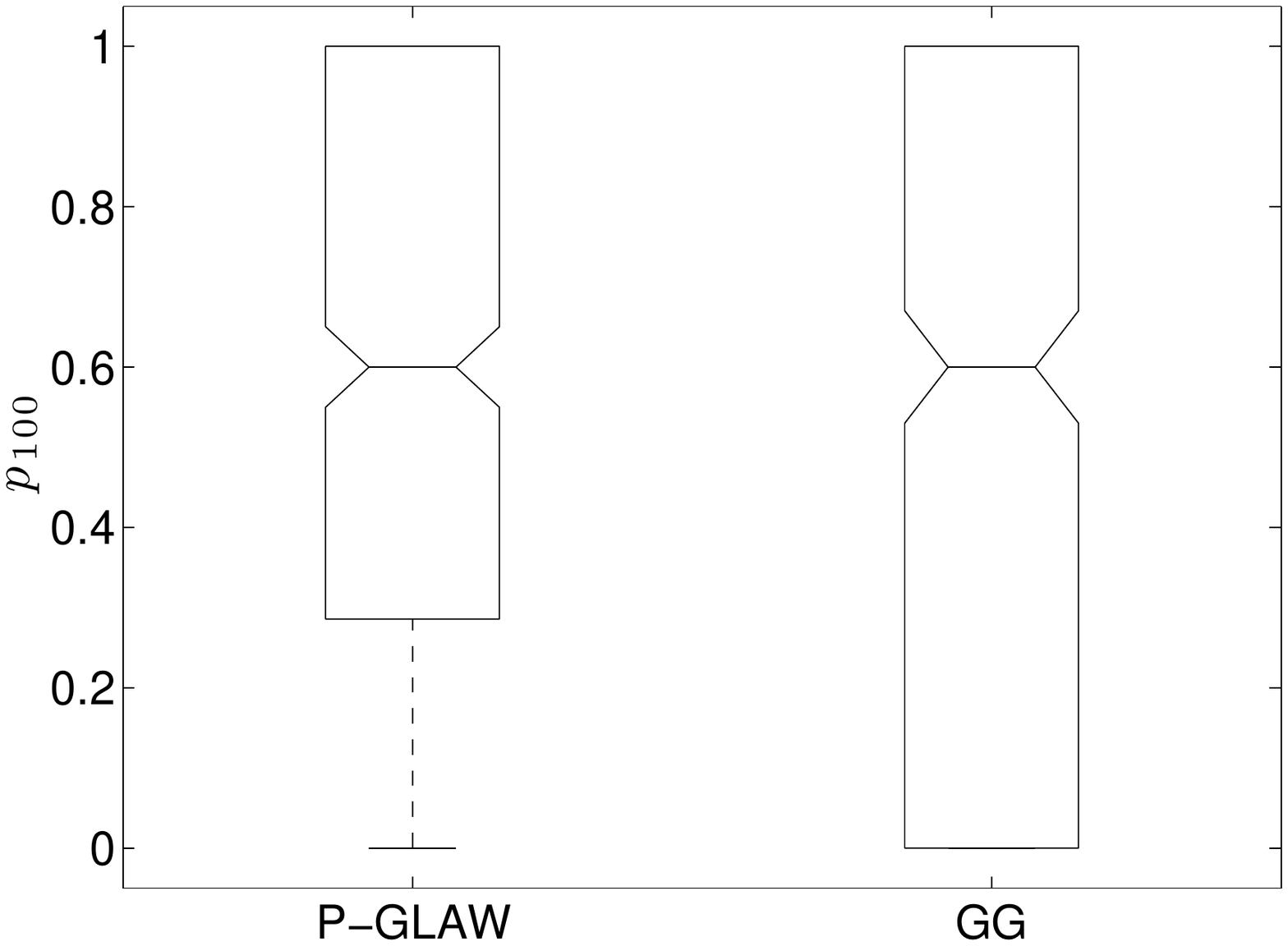}}
	\subfigure[(d) $S = 10; \delta_k = 0.001;$ random distbn]{\includegraphics[trim = 10mm 60mm 10mm 60mm, clip, scale=0.32]{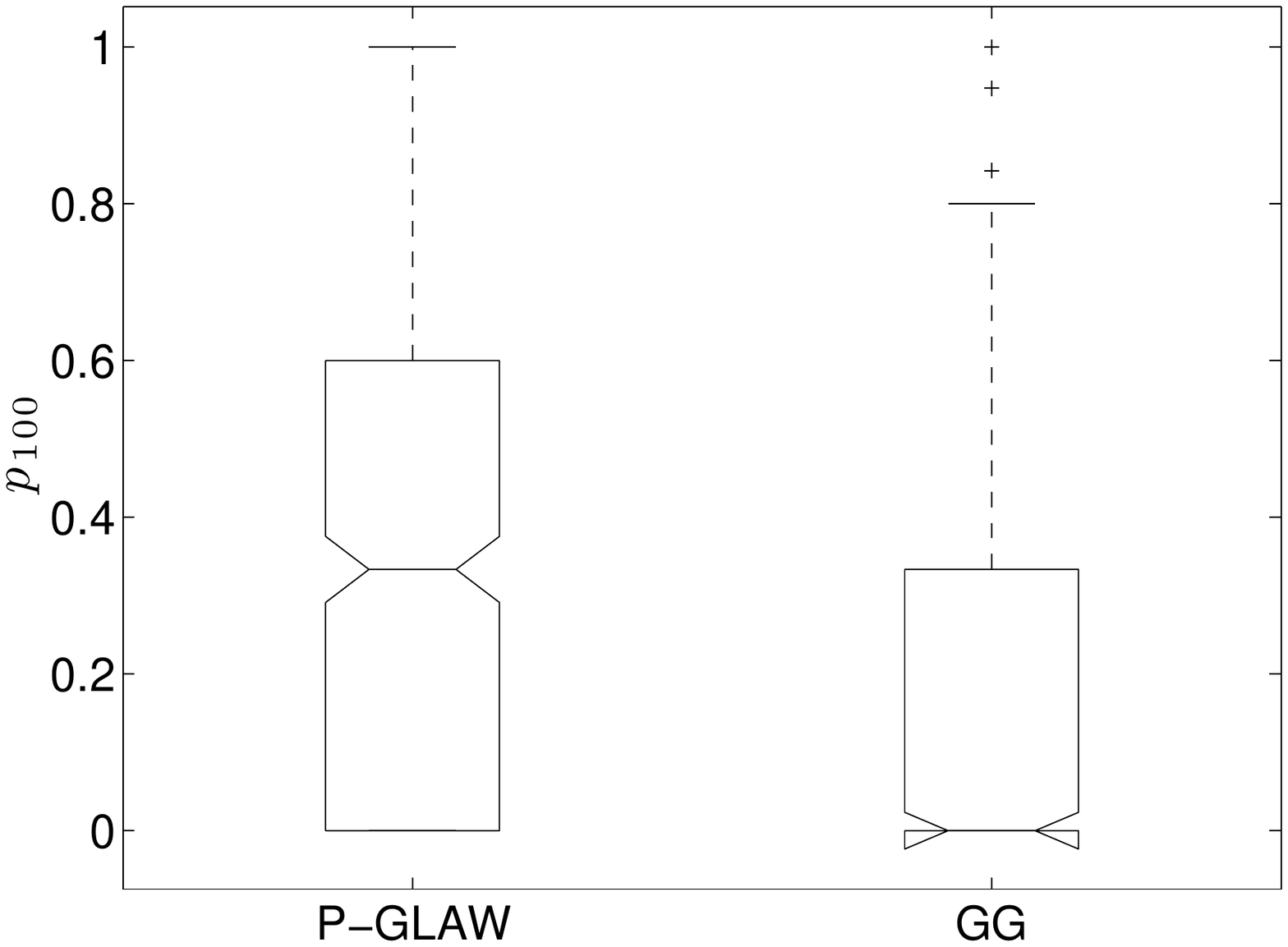}}
	\subfigure[(b) $S = 3; \delta_k = 0.005;$ random distbn]{\includegraphics[trim = 10mm 60mm 10mm 60mm, clip, scale=0.32]{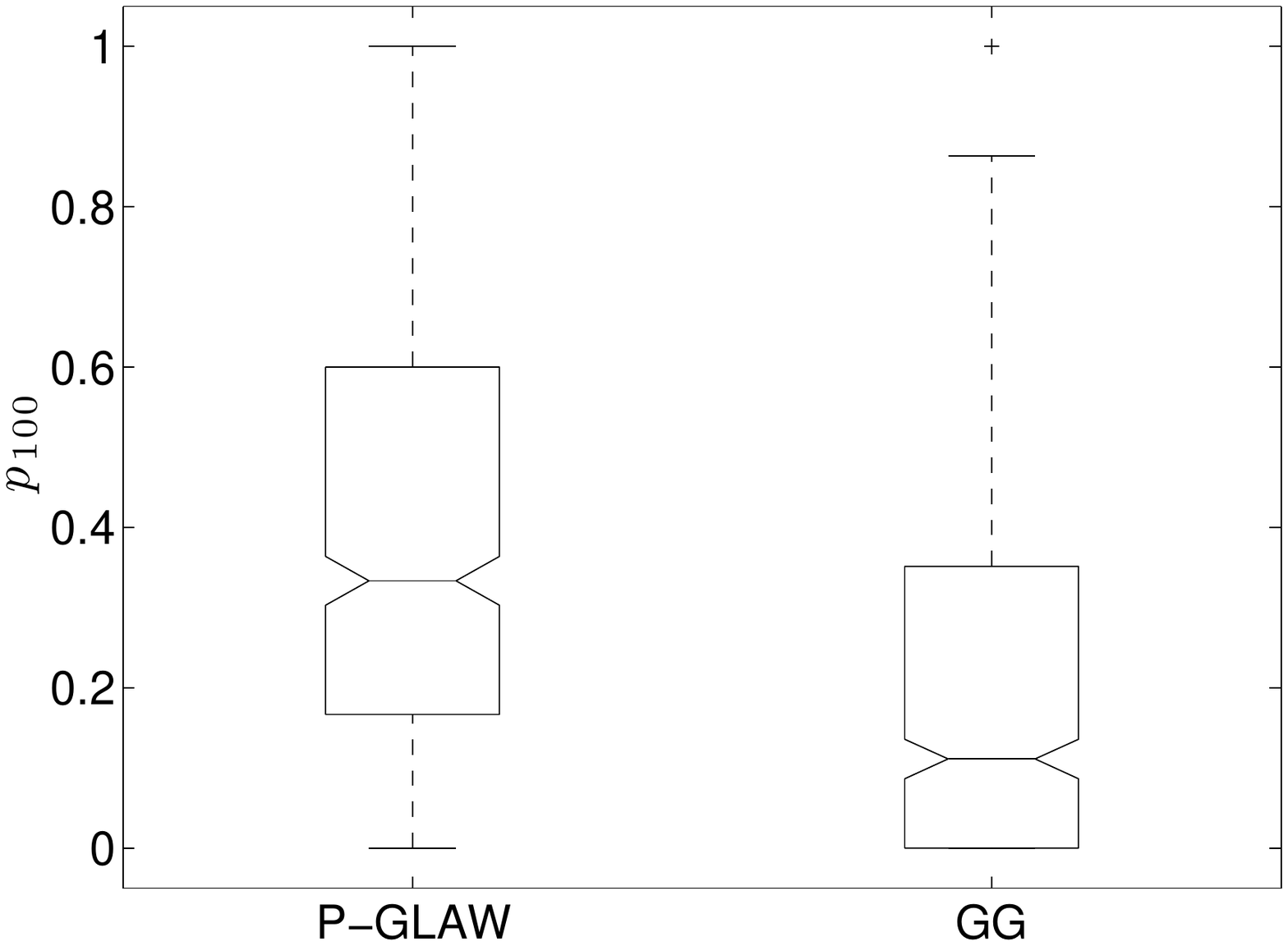}}
	\subfigure[(e) $S = 3; \delta_k = 0.001;$ random distbn]{\includegraphics[trim = 10mm 60mm 10mm 60mm, clip, scale=0.32]{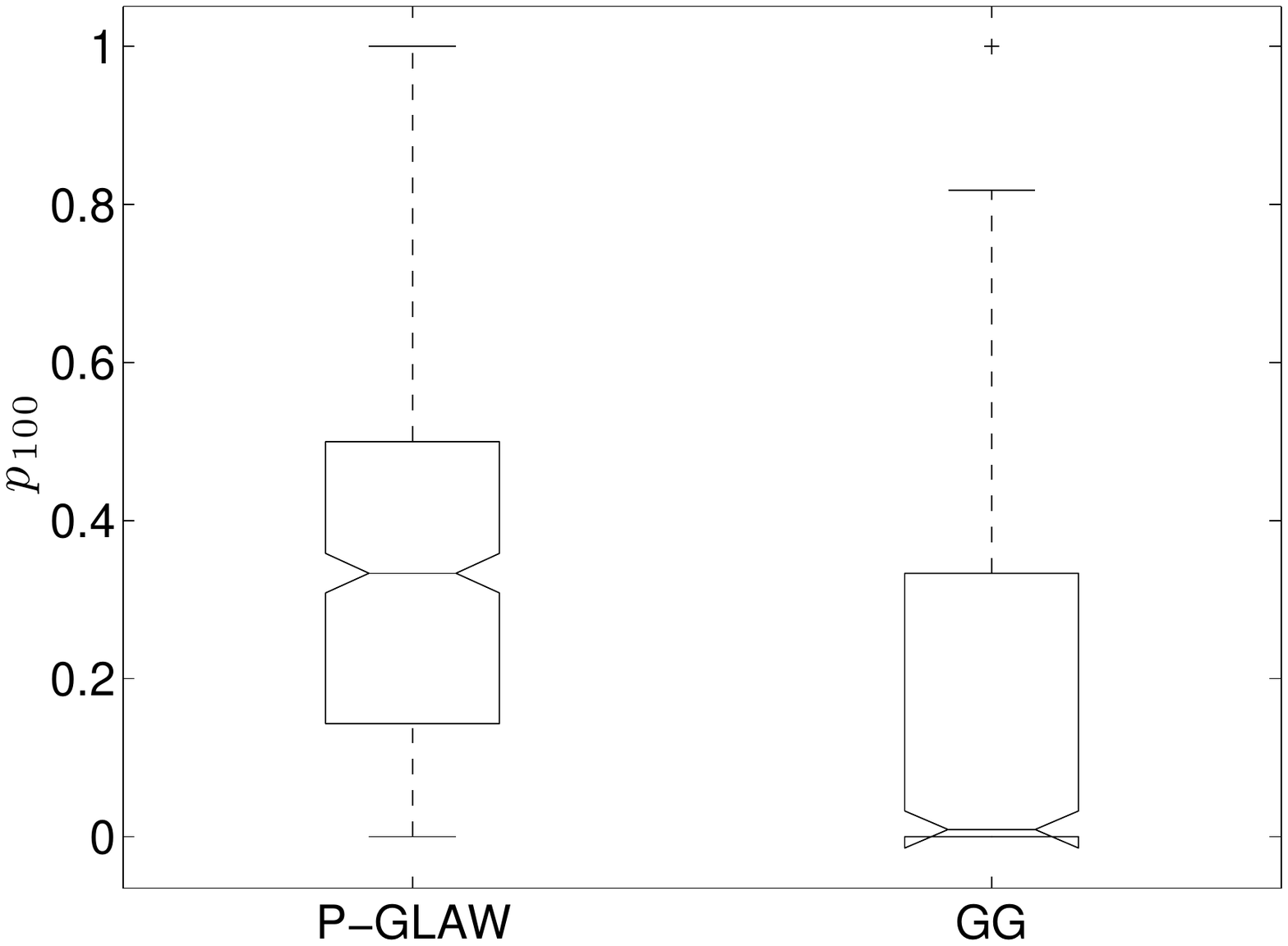}}
	\subfigure[(c) $S = 3; \delta_k = 0.005;$ single gene distbn]{\includegraphics[trim = 10mm 60mm 10mm 60mm, clip, scale=0.32]{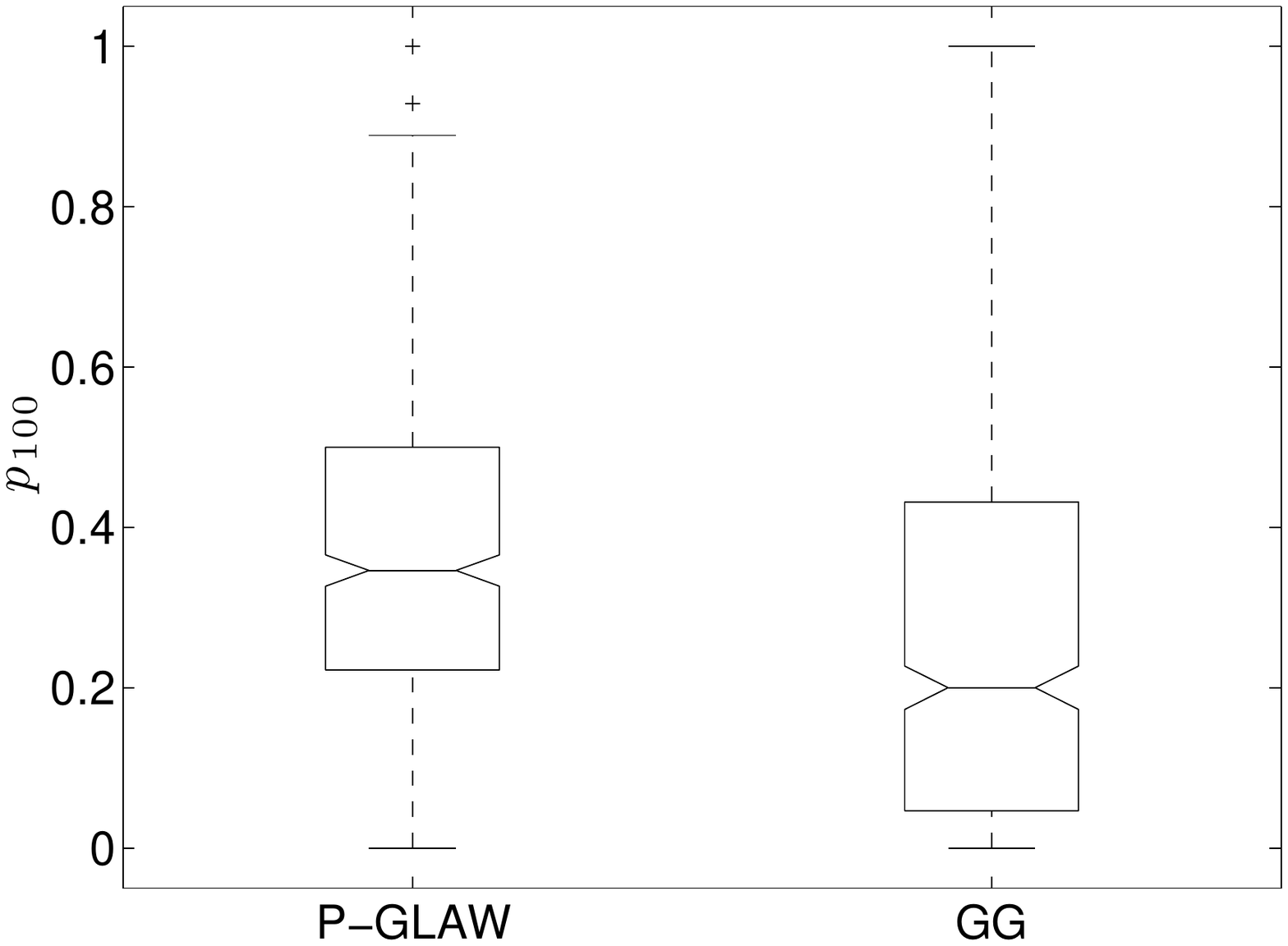}}
	\subfigure[(f) $S = 3; \delta_k = 0.001;$ single gene distbn]{\includegraphics[trim = 10mm 60mm 10mm 60mm, clip, scale=0.32]{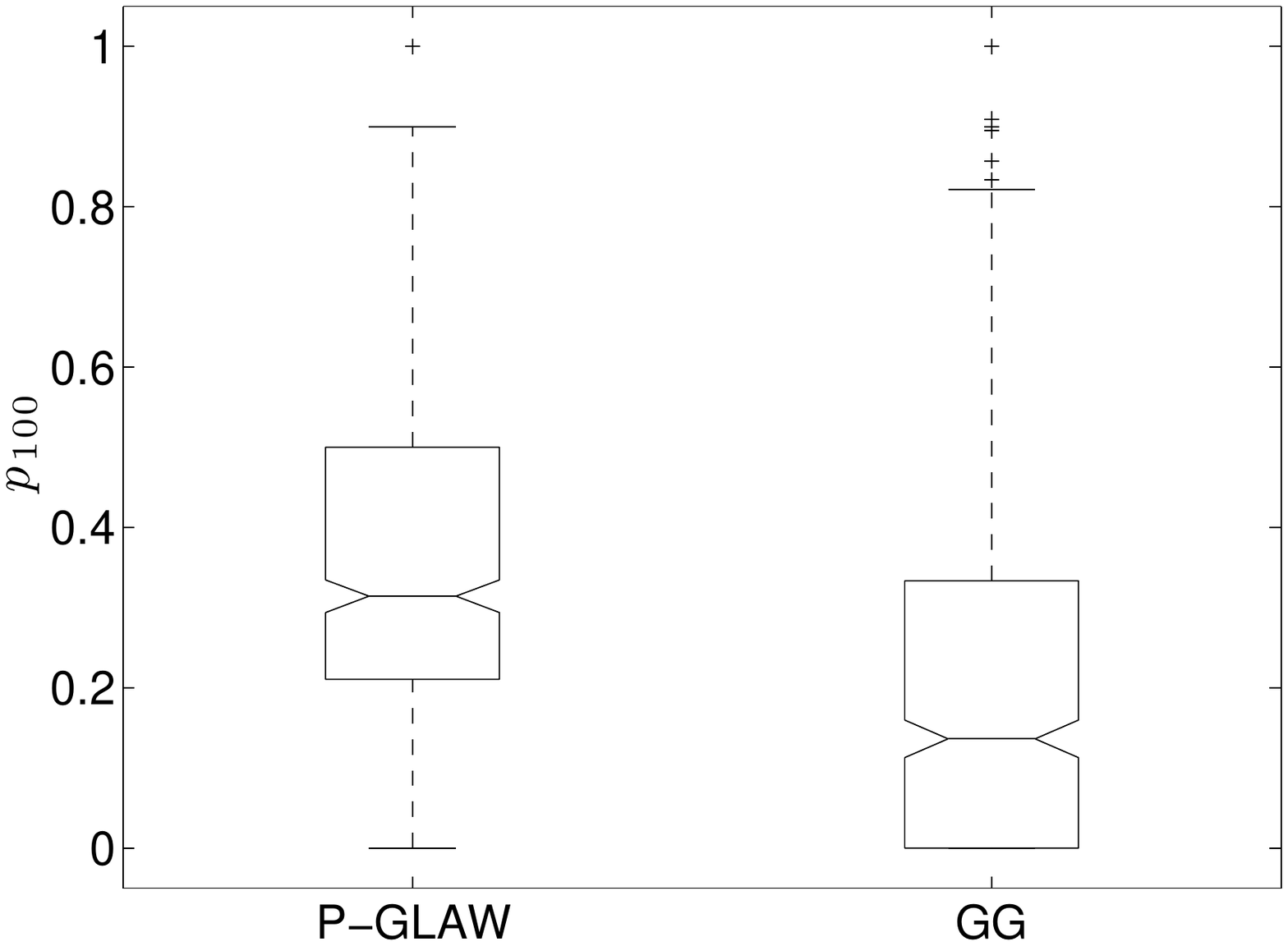}}	
	\caption{Box plots of distribution of ranking power, $p_{100}$, across $500$ simulations.  This is the proportion $|\mathcal{C}|^*_{100} / |\mathcal{C}|$ of causal pathways in $\mathcal{C}$ that are ranked in the top $100$ pathways.  Notches indicate $95\%$ confidence intervals for the true median.}
	\label{fig:p100}
	\end{center}
\end{figure}

% power100 boxplots
\begin{figure}
	\begin{center}
	\subfigure[(a) $S = 10; \delta_k = 0.005;$ random distbn]{\includegraphics[trim = 10mm 60mm 10mm 60mm, clip, scale=0.32]{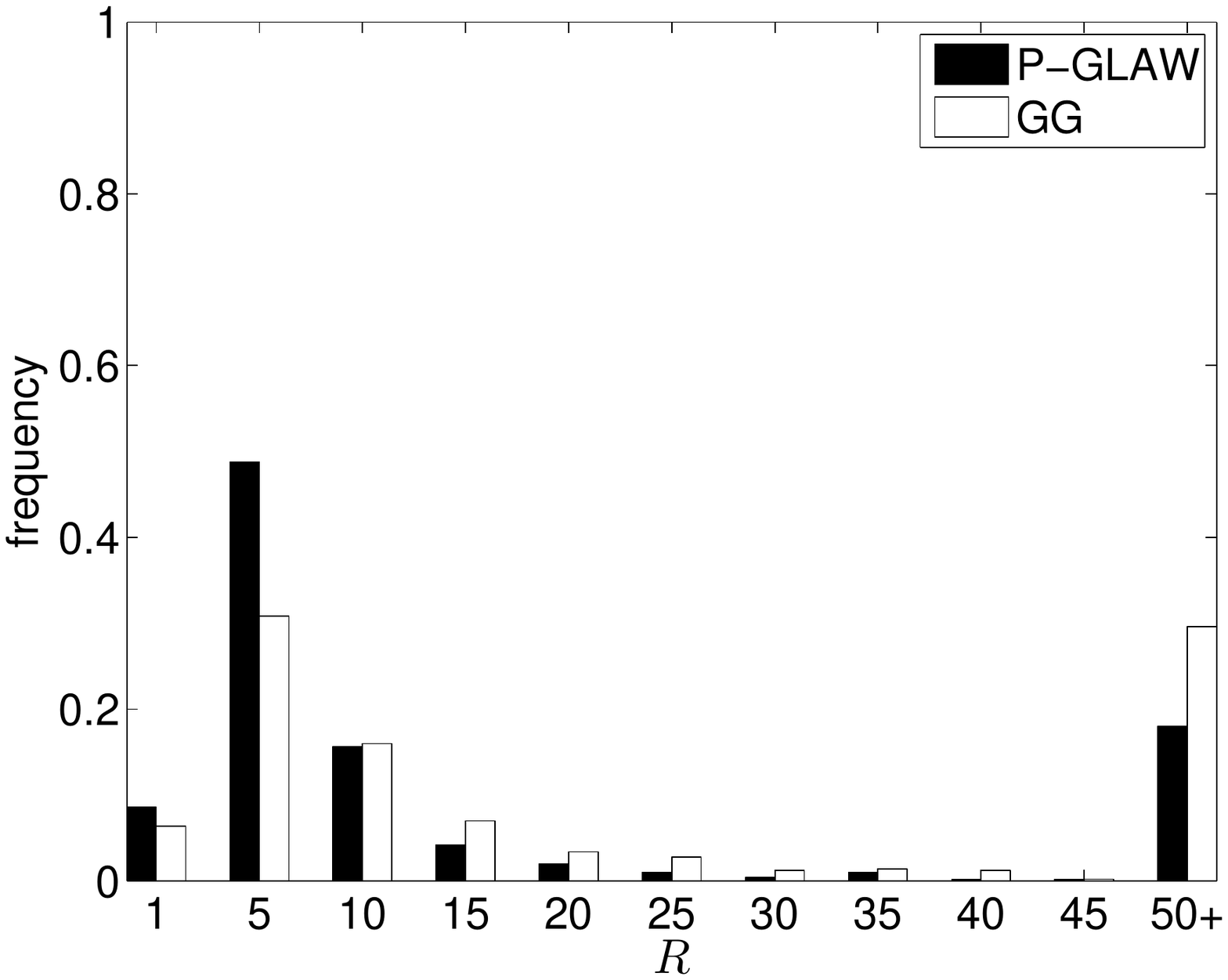}}
	\subfigure[(d) $S = 10; \delta_k = 0.001;$ random distbn]{\includegraphics[trim = 10mm 60mm 10mm 60mm, clip, scale=0.32]{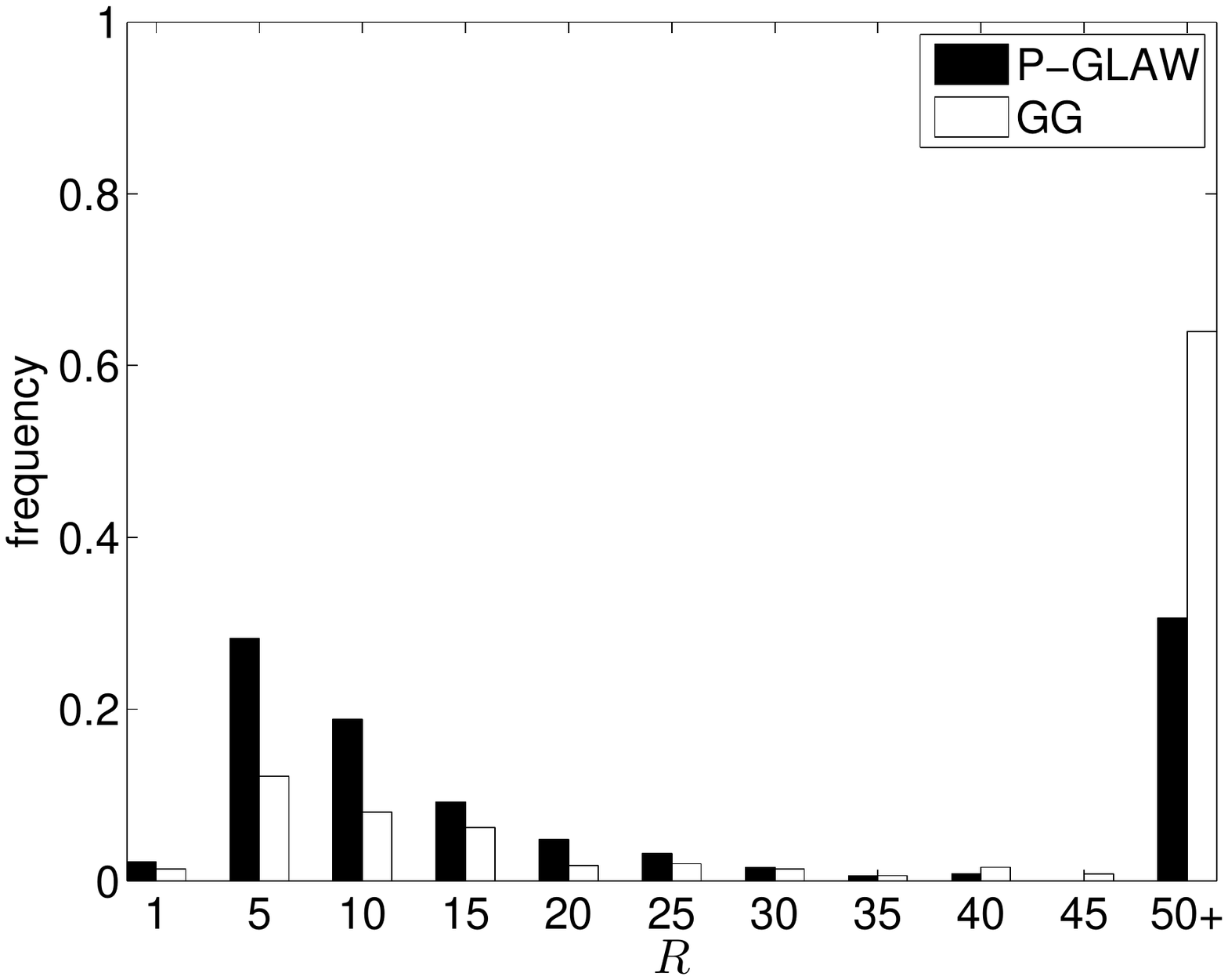}}
	\subfigure[(b) $S = 3; \delta_k = 0.005;$ random distbn]{\includegraphics[trim = 10mm 60mm 10mm 60mm, clip, scale=0.32]{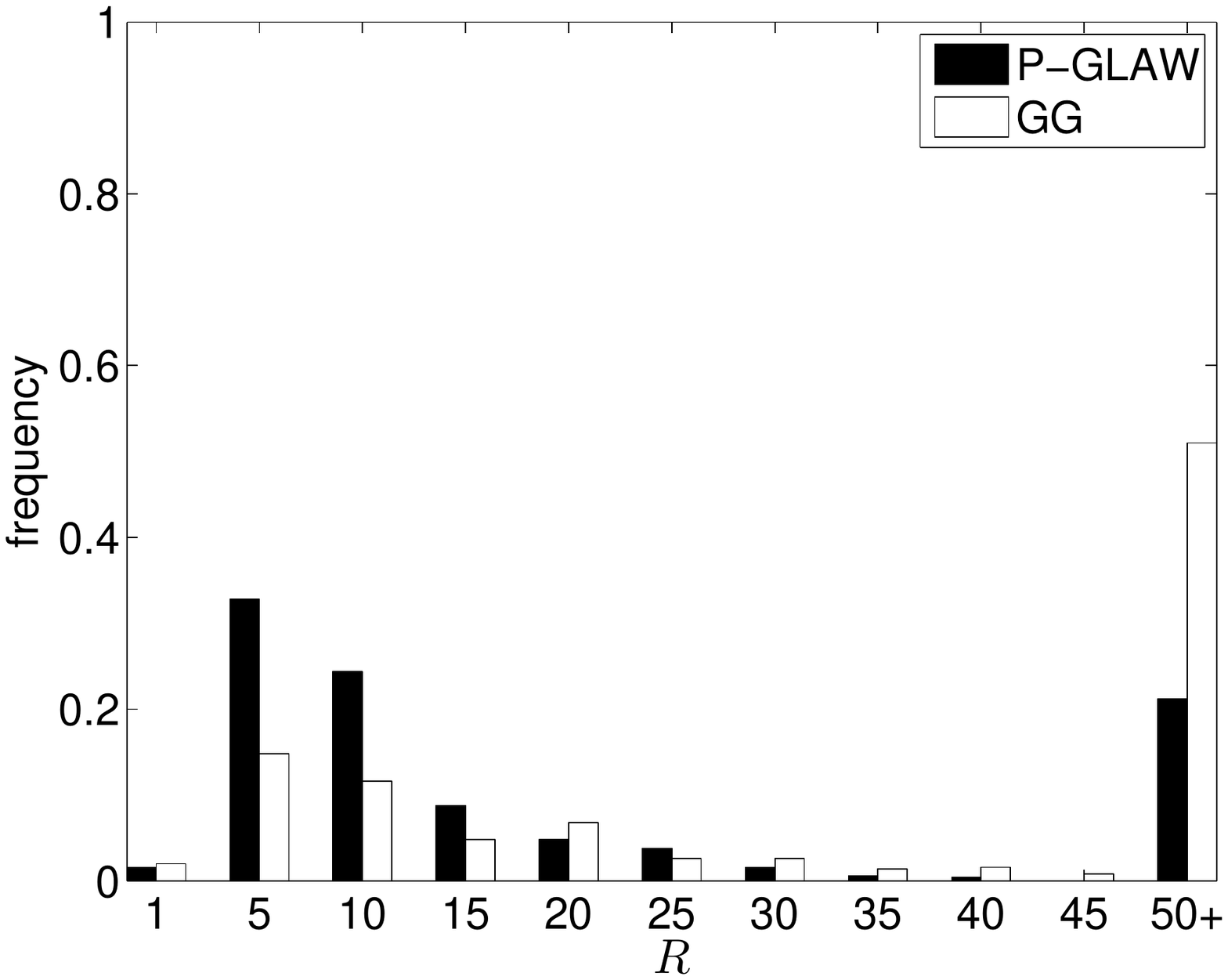}}
	\subfigure[(e) $S = 3; \delta_k = 0.001;$ random distbn]{\includegraphics[trim = 10mm 60mm 10mm 60mm, clip, scale=0.32]{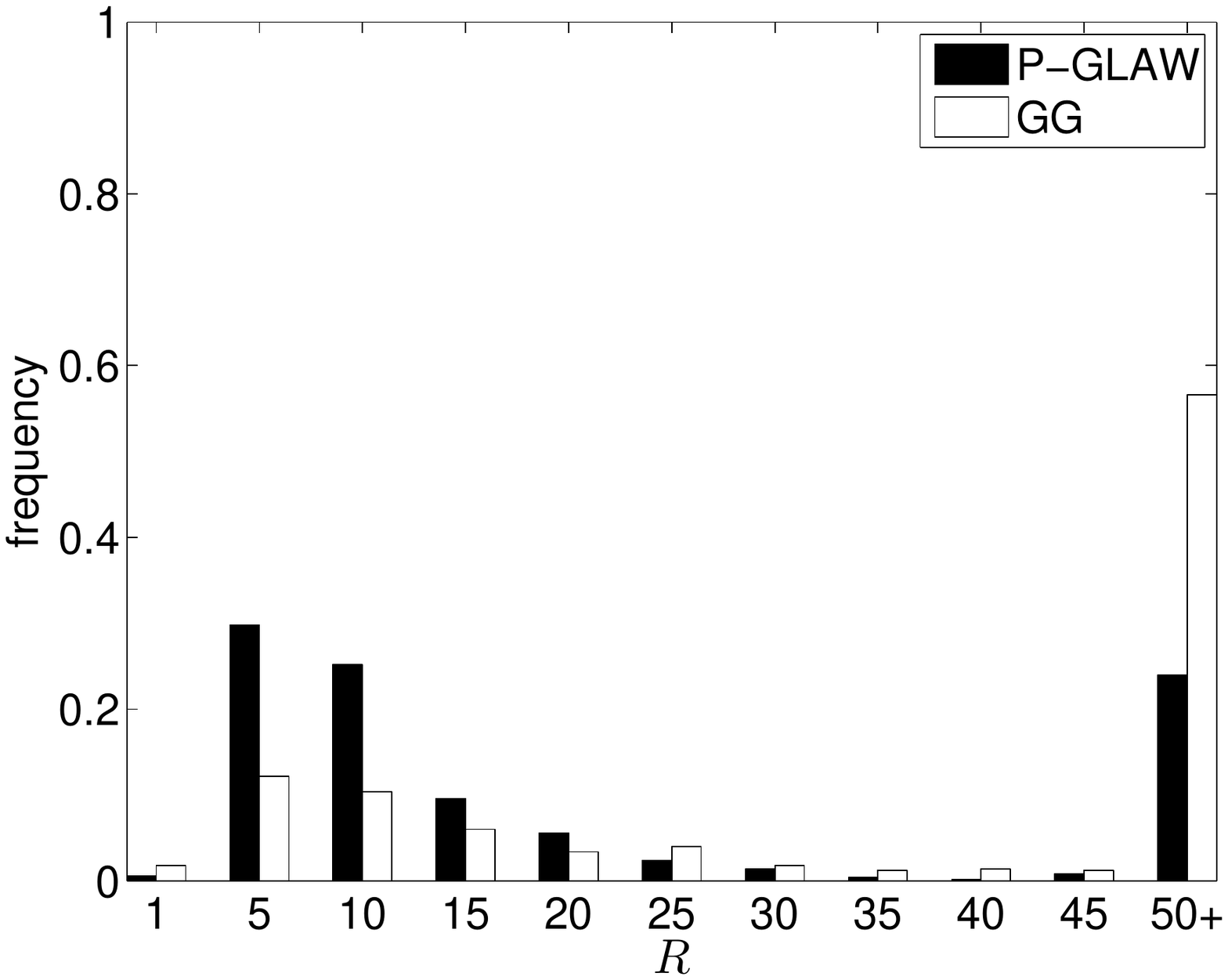}}
	\subfigure[(c) $S = 3; \delta_k = 0.005;$ single gene distbn]{\includegraphics[trim = 10mm 60mm 10mm 60mm, clip, scale=0.32]{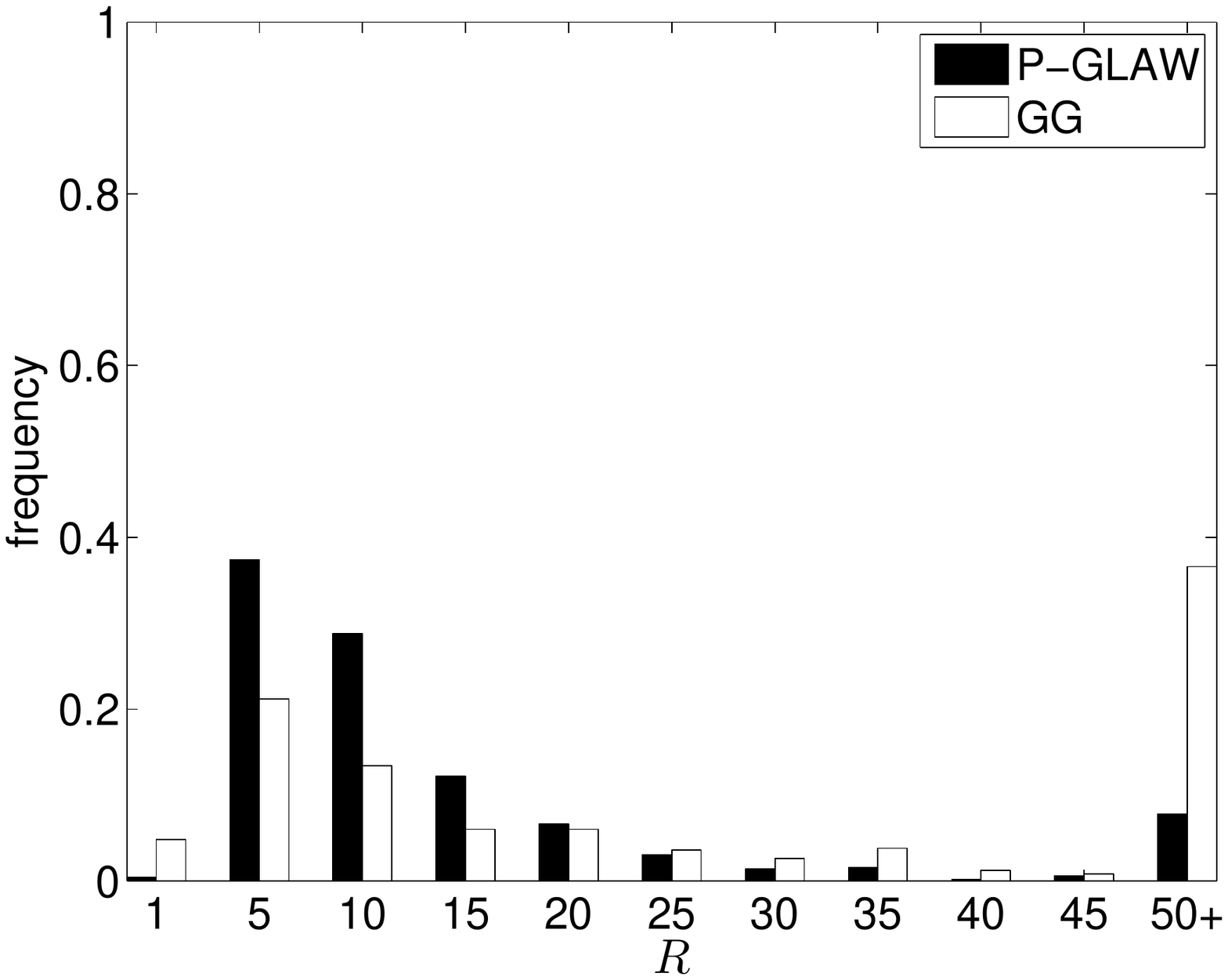}}
	\subfigure[(f) $S = 3; \delta_k = 0.001;$ single gene distbn]{\includegraphics[trim = 10mm 60mm 10mm 60mm, clip, scale=0.32]{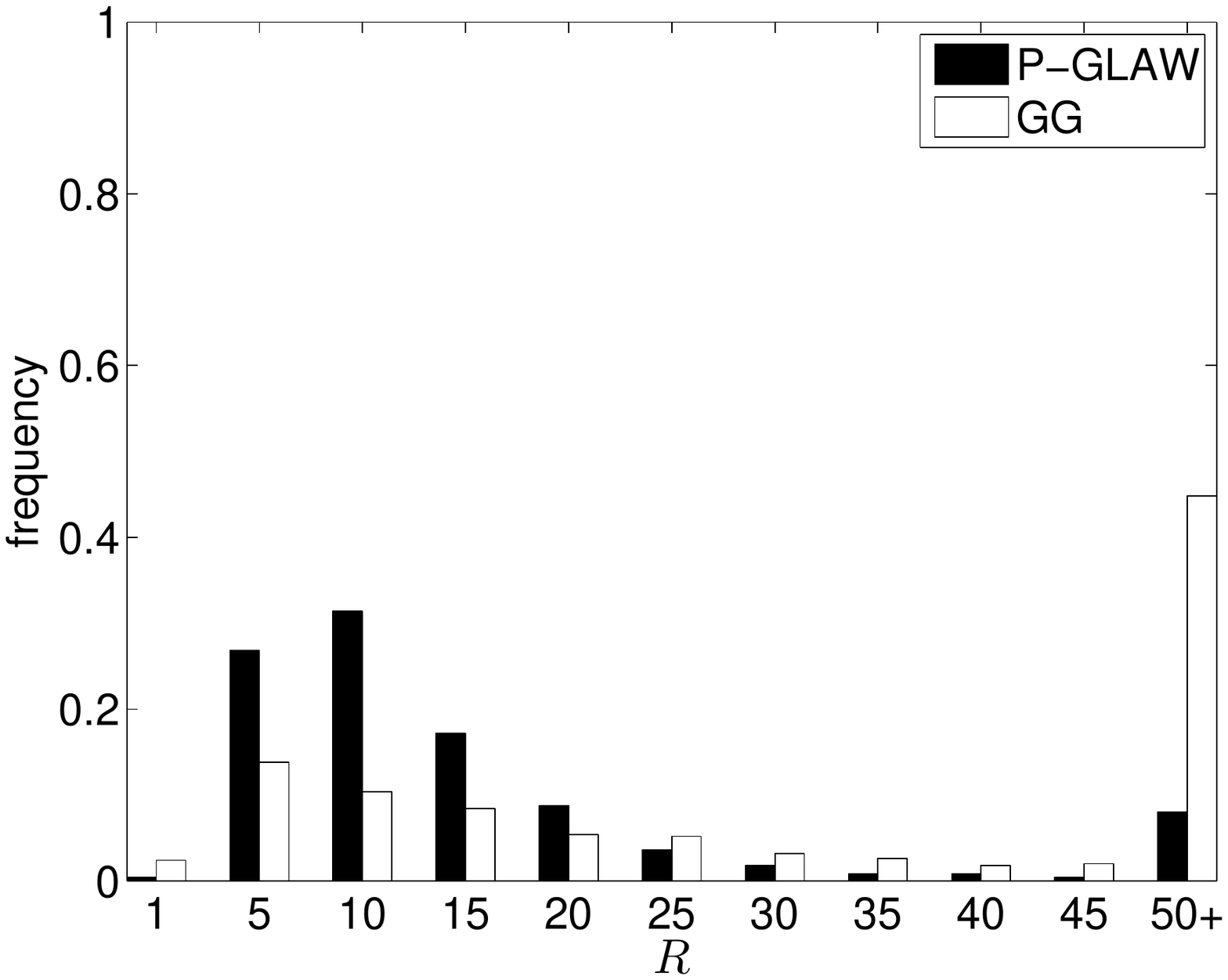}}	\caption{Distribution of the power-adjusted, normalised, weighted ranking score, $R$, across $500$ simulations.  The final `$50+$' column includes simulations for which no causal pathway was ranked in the top $100$, i.e.~$\mathcal{C}^*_{100} = \emptyset; R = 100$.}
	\label{fig:R100}
	\end{center}
\end{figure}

%%%%%%%%%%%%%%%%%%%%%%%%%%%%%%%%%%%%%%%%%%%%%%%%%%%%%%%%%%%%%%%%%%%%%%%%%%

\section{Discussion} \label{sec:Discussion}

We have developed a penalised regression-based strategy (P-GLAW) that exploits functional structure within genotypes to identify biological pathways associated with a continuous trait.  We use the group lasso, with all mapped SNPs and pathways in a single regression model, and use a novel combination of methods including a bias-adjusted group weighting scheme and bootstrap sampling, together with a number of speed ups designed to make the analysis of large scale datasets computationally feasible.  An important feature of our method is the need to accommodate the presence of overlapping pathways.  On the assumption that causal SNPs are enriched within a biological pathway, we find in a simulation study that our proposed method shows relative gains in both power and specificity across a range of scenarios, compared with an alternative pathways method (GG), based on univariate SNP statistics, that we use as a benchmark.  We believe this is the first such study evaluating GL performance using real SNP and pathway data across a range of realistic scenarios.

One key motivation for a pathways-based approach is the desire to harness the joint effects of those SNPs or genes with relatively small effect size, that typically fail to achieve genome-wide significance in GWAS \citep{Baranzini2009}.  We hypothesise that the advantages inherent in a multivariate approach to modelling SNP effects will increase power to detect these, and in our simulation study we therefore focus on scenarios with causal SNPs that exhibit effect sizes at or below the limits of those found in recent GWAS.  To evaluate the performance of each method considered here, we devise three separate ranking metrics, each of which measures a different aspect of ranking performance.  

One factor affecting power is the `genetic architecture' of the disease in question, that is the number and distribution of SNP effects across causal pathways \citep{Wang2010}.  For example, causal SNPs may be distributed across many genes in a pathway, or restricted to a single gene.  Since PGAS methods vary in the way that they combine the effects of individual SNPs, the specific genetic architecture is expected to impact power for different methods in different ways \citep{Wang2009,Holmans2009}.  GG uses genes scores corresponding to the most significant SNP associated with a gene to establish pathway significance.  This has the advantage of reducing redundant information arising from SNPs in LD with a causal SNP within a single gene, but may lead to reduced power where causal variants reside in distinct LD blocks within a gene \citep{Wang2007}.  An important, related factor that we find has received little attention is the issue of overlapping pathways, and the consequent effect on PGAS performance.  The precise distribution of causal SNPs with respect to genes that overlap multiple pathways will affect the number of pathways that are considered to be `causal', and we expect this to affect ranking performance for different methods in different ways.  To explore these issues, we investigate a variety of different genetic architectures, in which we vary both the number and distribution of causal SNPs with respect to pathways and genes.

In general, we find that P-GLAW performs well across the range of causal SNP distributions and effect sizes considered.  Additionally, our method is able to consistently outperform the benchmark (GG).  GG performance at the smaller effect size is particularly weak, so that P-GLAW shows the largest gains in relative performance here.

An insight into some of those factors affecting ranking performance is afforded by considering some of the ranking measures in more detail.  Starting with the highest ranking causal pathway measure ($r_{k_1})$, as expected we find that this decreases for each scenario at the smaller effect size.  However, at the smaller effect size this measure is observed to increase for both methods as the number of causal SNPs is decreased, markedly so when the reduced number of causal SNPs are concentrated in a single gene.  Since the effect size for each causal SNP is held constant, this seems counterintuitive, since the pathway `signal' is reduced when there are fewer causal SNPs.  In addition, for the reasons described above, for GG this signal may be further reduced where causal SNPs reside within a single gene.  The explanation is likely to be that the effective number of causal pathways tends to increase as the number of causal SNPs is reduced, increasing the probability that a single causal pathway is ranked high.  The number of causal pathways is even larger when causal SNPs are concentrated in a single gene (see Fig.~\ref{fig:nCausalPaths}).  Where the pathway signal is highest (scenario (a)), both methods tend to rank a high proportion of causal pathways in the top 100 (high $p_{100}$), although the proportion of MC simulations in which GG fails to rank any causal pathways (that is the proportion of simulations with $p_{100} = 0$) is relatively high.  On this measure of ranking power, GG performs relatively poorly across all other scenarios, particularly at the smaller effect size.  Interestingly, P-GLAW is relatively insensitive to variation in the number and distribution of SNPs within causal pathways, as might be expected from the smoothing properties of the GL $\ell_2$ penalty, which ensures that all SNPs within a selected pathway are retained in the model \citep{Zhou2010}.

The need to account for factors such as variation in LD, gene and pathway size is a feature common to all PGAS methods.  A range of approaches, often used in combination, have been proposed to correct for these biasing factors, including the use of gene scores that summarise SNP statistics \citep{Holmans2009}, and permutation of phenotypes \citep{Wang2009}.  Dimensionality reduction techniques have also been advocated for the control of redundant information \citep{Chen2010,Zhu2011,Ballard2010}.  For P-GLAW, we propose a method that adjusts the distribution of pathway weights according to the observed bias in pathway selection frequencies across multiple MC simulations under the null.  We find in a simulation study that our proposed bias correction method does substantially increase power and specificity, indicating that pathway selection bias is decreased.  One potential disadvantage of our approach is that it takes no account of variation in biasing factors within a pathway.  It would be interesting to compare the relative merits of our approach against alternative bias-reduction methods, for example the use of within-pathway dimensionality reduction.  However, we consider the retention of all SNPs in the regression model to be a potentially attractive feature of our approach, as it affords the possibility of the simultaneous identification of causal SNPs driving pathway selection, and we are currently pursuing this as an extension to the present model.

In situations where predictors, or groups of predictors are correlated, both the lasso and group lasso can demonstrate problems with consistency, that is they are unable to constently identify the true set of causal predictors or groups \citep{Yu2006,Bach2008,Chatterjee2011}.  Despite this, we have demonstrated that in a finite sample, our bootstrap sampling approach performs well, and this has been borne out elsewhere \citep{Meinshausen2010}.  We are however pursuing alternative methods for the ranking of pathways, using different ranking strategies.

We pay considerable attention to the need to develop fast algorithms for solving the GL, a problem that is particularly acute when using regression models with GWAS data.  Using a combination of techniques, we establish a GL estimation algorithm that can quickly solve the GL using whole genome data.  However, the very large number of simulations and scenarios considered in our simulation study, and the relatively slow performance of the benchmark method mean that we restrict the analysis to mapped SNPs from a single chromosome.\footnote{Python code for mapping SNPs to pathways, and for analysing SNP data using PGLAW is available on request.}

We note that phenotypes in our simulation study are generated under an additive linear model.  The assumption of additive linear SNP effects is built into both the P-GLAW and GG models, in the former through the SNP allele codings in the genotype design matrix, and in the latter through the particular model used to generate the univariate SNP scores, although for both methods alternative models can easily be accommodated.  

In our simulation study we account for variation in the size and distribution of causal SNP minor allele frequencies through the use of MC simulations, but we expect that such variation is likely to impact model performance, and this is something that warrants further exploration.

As with all PGAS methods, we note that results are dependent on the choice of pathways database, and will inevitably reflect biases due for example to the increased number of annotations for genes implicated in particular disease etiologies \citep{Elbers2009,Cantor2010}.  Results are also subject to bias resulting from SNP to gene mapping strategies.  For example, SNP to gene mapping distances will affect the number of unmapped SNPs falling within gene `deserts' \citep{Eleftherohorinou2009}, SNPs will map to relatively large numbers of genes in gene rich areas of the genome, and the mapping of a SNP to its closest gene may obscure a true functional relationships with a more distant gene \citep{Wang2009}.

Finally, we note that our method can be easily adapted to accommodate other ways of grouping SNP data, for exampling using protein interaction networks \citep{Wu2010a}, or GO and other ontologies \citep{Jensen2010}.
%%%%%%%%%%%%%%%%%%%%%%%%%%%%%%%%%
\setcounter{secnumdepth}{-1}
\section{Appendix \tab Line search over $\lambda$} \label{appendix:lambdaLineSearch}
We wish to tune $\lambda$ so as to select a minimum $M$ pathways at each subsample.  To do this we perform a line search over $\lambda$.  This procedure is described in box 3. \\
%
%%%%%%%%%%%%%%%%%%%%%%%%%%%%%%%%%
% BOX FOR GOLDEN SECTION SEARCH OVER LAMBDA USING ACTIVE SET

\vspace{3pt}
\noindent\textbf{Box 3} Line search procedure for tuning $\lambda$ to select $M \ge M_{min}$ pathways\\
\tiny\normalsize
\begin{enumerate}
\item
	Set $\lambda_{max} = \min_{\lambda}: || \mathbf{X}_l^T \mathbf{y} ||_2 \le \lambda w_l$ (from \eqref{eq:lambdaMax}) and $\alpha = 0.8^{\dagger}$
	\item
	Let $\lambda = \alpha \lambda_{max}$
	\item
	Form the active set, $\mathcal{A} = \{ m \in \mathcal{G} : || \mathbf{X}_m^T \mathbf{y} ||_2 \le \lambda w_m \}$
	\item
	Let $M = | \mathcal{A} | $.  If $M < M_{min}$ skip to step 6.$^{\ddagger}$
	\item
	Solve the GL estimation at $\lambda$ using the active set $\mathcal{A}$, as described in box \ref{box:activeSetSingleLambda} (starting at box \ref{box:activeSetSingleLambda}, step 2.)
	\begin{tabbing}
	kdk\=kdk\=kdk\=kdk\=kdk\=kdk\=kdk\=kdk\=kdk\kill % set tabs
	Let the solution be $\hat{\boldsymbol{\beta}}$, with final active set $\mathcal{A}$\\
	$\mathcal{S}(\lambda) =\{l \in \mathcal{G}: || \hat{\boldsymbol{\beta}}_l || > \mathbf{0}\}$ \>\>\>\>\>\>\>\> (the set of selected pathways)\\
	$M = |\mathcal{S}(\lambda)|$												\>\>\>\>\>\>\>\> (the number of selected pathways)
	\end{tabbing}
	\item
	\begin{tabbing}
	kdk\=kdk\=kdk\=kdk\=kdk\=kdk\=kdk\kill % set tabs
	if $M \ge M_{min}$\\
		\> $\hat{\boldsymbol{\beta}}$ is the full solution\\
		\> STOP\\
	else\\
		\> $\lambda_{max} = \lambda$	\>\>\>\> (need to decrease $\lambda$)\\
	end
	\end{tabbing}
	\item
	Go to step 2.
\end{enumerate}
{\small $^\dagger$ The value of $\alpha$ is chosen for computational convenience.  A value close to $1$ ensures that $\lambda$ values stay close to $1$, so that as few pathways are selected by the model as possible, thus speeding up the estimation.  However, a value too close to $1$ means that the decrease in $\lambda$ at each iteration is small, meaning that many iterations may have to be performed before $M$ reaches the desired range.\\
$^{\ddagger}$ This step is introduced for computational efficiency, since if $| \mathcal{A} | < M_{min}$ there is no prospect of selecting enough groups}
\vspace{6pt}

%-----------------------------------------------------------
\bibliographystyle{BEPress}
\bibliography{library}	% location of .bib file
%-----------------------------------------------------------
\end{document}